\documentclass[twocolumn,amsmath,amssymb,prd,floatfix,nofootinbib]{revtex4}

\usepackage{graphicx}
\usepackage{dcolumn}
\usepackage[varg]{txfonts}
\usepackage{bm}

\usepackage{color}

\begin{document}

\title{Velocity bias and the nonlinear perturbation theory of peaks}

\author{Takahiko Matsubara} \email{tmats@post.kek.jp}
\affiliation{%
  Institute of Particle and Nuclear Studies, High Energy Accelerator
  Research Organization (KEK), Oho 1-1, Tsukuba, Ibaraki 305-0801, Japan
}%
\affiliation{%
  The Graduate University for Advanced Studies (SOKENDAI), Tsukuba,
  Ibaraki 305-0801, Japan
}%

\date{\today}

\begin{abstract}
  The biasing in the large-scale structure of the universe is a
  crucial problem in cosmological applications. The peaks model of
  biasing predicts a linear velocity bias of halos, which is not
  present in a simple model of local bias. We investigate the origin
  of the velocity bias in the peaks model from the viewpoint of the
  integrated perturbation theory, which is a nonlinear perturbation
  theory in the presence of general Lagrangian bias. The presence of
  the velocity bias in the peaks model is a consequence of the ``flat
  constraint,'' $\bm{\nabla}\delta = 0$; i.e., all the first spatial
  derivatives should vanish at the locations of peaks. We show that
  the velocity bias in the peaks model is systematically derived in
  the framework of the integrated perturbation theory, and then
  develop a formal theory to perturbatively trace the nonlinear
  evolution of biased objects with the flat constraint. A formula for
  the nonlinear velocity dispersion of peaks with the one-loop
  approximation is also derived.
\end{abstract}

\maketitle


\section{\label{sec:Intro}
Introduction
}

The large-scale structure (LSS) of the universe is one of the main
sources of cosmological information. The statistical properties of LSS
constrain the physics of the early universe as well as the nature of
dark matter, dark energy, etc. The large-scale structure is mainly
probed by observable objects such as galaxies, quasars, and other
astronomical objects. The spatial distribution of objects is not the
same as that of dark matter; the observable objects are biased tracers
of matter distributions. The precise relation between the objects and
the dark matter is complicated because of, e.g., nonlinear,
nongravitational physics, etc. On large scales where linear theory is
applicable, it is a common practice to simply apply a linear bias.
However, the linear bias is too simplistic to extract maximum
information from the LSS which experiences nonlinear effects of
gravitational evolution.

Modeling the nonlinear biasing is a theoretically nontrivial problem
in general. One of the natural formulations of the bias modeling is
provided by the peak approach \cite{BBKS}. In this approach, the
density peaks of the initial density field are identified as formation
sites of astronomical objects. The peaks model predicts a linear
velocity bias of halos \cite{Des08,DS10,DCSS10,Bal15}. Unlike the
prediction of the coupled-fluids approximation for the coevolution of
dark matter and halos \cite{Fry96,TP98,CSS12}, the velocity bias in
the peaks remains constant in time. This property is explained by the
modification of the halo momentum conservation equation in the
presence of peak constraints \cite{BDKR14}.

The time evolution of the statistical properties of peaks with the
velocity bias has been investigated mainly by applying the Zel'dovich
approximation \cite{Zel70}. The Zel'dovich approximation corresponds
to the first-order approximation in terms of the Lagrangian
perturbation theory \cite{Buc89,Mou91,Buc92}. A Lagrangian
perturbation theory in the presence of general bias is given by the
integrated perturbation theory (iPT) \cite{Mat11,Mat14}. In this
formalism, any form of the bias model in Lagrangian space can be taken
into account \cite{MD16}, including the halo bias \cite{MW96,MJW97},
peaks model \cite{BBKS,LMD15}, and excursion set peaks
\cite{AP90,PS12}. Therefore, the dynamical evolution of peaks can be
described by iPT including higher-order effects of perturbations
beyond the Zel'dovich approximation.

In this paper, we apply the iPT to investigate the properties of peak
evolution, focusing on the origin of velocity bias in the peaks model.
We see how the velocity bias appears in the formalism of iPT with peak
bias and derived expressions of two- and three-point propagators which
are ingredients to predict the one-loop approximation of the power
spectrum. In the limit of the Zel'dovich approximation, previously
known results of the velocity bias of peaks are reproduced. We give an
analytic expression of the velocity dispersion of peaks in the
one-loop approximation. Then we consider the velocity bias from a more
formal point of view and show that the velocity bias originates from
the ``flat constraint'' in the peaks model. The flat constraint is the
condition that the first spatial derivatives of the field should
vanish at the location of peaks. That is, any bias model with the flat
constraint, such as the peaks model and excursion set peaks, should
predict the velocity bias. We consider a resummation theory of the
flat constraint and derive resummed propagators of density and
velocity.

In this way, the origin and properties of velocity bias in the peaks
model are theoretically investigated in this work. Our paper is
organized as follows. In Sec.~\ref{sec:LinearVelocityBias}, the
presence of linear velocity bias is derived in the framework of iPT.
The two- and three-point propagators of density and velocity, which
are ingredients of predicting the power spectrum, are derived. In
Sec.~\ref{sec:VelocityDispersion}, the one-loop expression of velocity
dispersion is derived. The resulting expression is given by
combinations of one-dimensional fast Fourier transforms (FFT). In
Sec.~\ref{sec:General}, a formal aspect of the higher-order
perturbation theory in the presence of velocity bias is investigated.
Conclusions are summarized in Sec.~\ref{sec:Conclusion}. Detailed
derivations of necessary functions are given in
Appendix~\ref{app:IntegralSummation}.

\section{\label{sec:LinearVelocityBias}
Velocity bias in the peaks
}

In this section, we show how the velocity bias is derived in the iPT
of lower-order perturbations.

\subsection{\label{subsec:VelocityField}
Velocity of peaks and iPT
}

In the bias models such as the peaks model, the biased objects are
discretely distributed in space. In this case, the velocity field of
biased objects is not a continuous field. There is not any value of
velocity at every location where discrete objects do not reside;
the velocities of the discrete objects are defined only at the
locations of objects. Therefore, a natural quantity to describe the
velocity of discrete objects are the number-weighted velocity field,
or the momentum,
\begin{equation}
  \bm{j}_X(\bm{x}) = [1 +
  \delta_X(\bm{x})]\bm{v}(\bm{x}), 
  \label{eq:2-1-01}
\end{equation}
where $\delta_X(\bm{x})$ is the number density contrast of objects $X$
at a location $\bm{x}$, and $\bm{v}(\bm{x})$ is the peculiar velocity
at the same location. Throughout this paper, time dependencies are
omitted in the arguments of various variables, although they actually
depend on time. 

For the Lagrangian-space counterpart, we have
\begin{equation}
  \bm{j}_X^\mathrm{L}(\bm{q}) = [1 + \delta_X^\mathrm{L}(\bm{q})]
  \bm{v}^\mathrm{L}(\bm{q}),
  \label{eq:2-1-02}
\end{equation}
where the Lagrangian velocity field $\bm{v}^\mathrm{L}$ is defined by
$\bm{v}^\mathrm{L}(\bm{q}) \equiv \bm{v}(\bm{x})$, where
$\bm{x}=\bm{q} + \bm{\varPsi}(\bm{q})$ and $\bm{\varPsi}(\bm{q})$ is
the displacement field. The Eulerian and Lagrangian momenta are
related by
\begin{equation}
  \bm{j}_X(\bm{x}) =
  \int d^3\!q\, \bm{j}_X^\mathrm{L}(\bm{q})\,
  \delta_\mathrm{D}^3[\bm{x}-\bm{q}-\bm{\varPsi}(\bm{q})].
  \label{eq:2-1-03}
\end{equation}
The above relation is derived by noting a relation
$[1 + \delta_x(\bm{x})]d^3x = [1+\delta^\mathrm{L}_X(\bm{q})]d^3q$.
The Lagrangian velocity $\bm{v}^\mathrm{L}$ in Eq.~(\ref{eq:2-1-02})
is given by
\begin{equation}
  \bm{v}^\mathrm{L}(\bm{q})
  = a \dot{\bm{\varPsi}}(\bm{q}),
  \label{eq:2-1-04}
\end{equation}
where $a$ is the scale factor, and the dot denotes a derivative with
respect to time,
$\dot{\bm{\varPsi}} = \partial\bm{\varPsi}/\partial t$. The momentum
$\bm{j}_X(\bm{x})$ is nonzero only at the locations of discrete
objects, so Eq.~(\ref{eq:2-1-04}) is evaluated only at locations of
discrete objects as well (at locations of peaks, in the case of the
peaks model). Thus the velocity of discrete objects is {\em
  statistically} biased with respect to the velocity of mass, even
when one assumes that velocities of matter and objects are the same at
locations of discrete objects.

Throughout this paper, it is assumed that the peculiar velocity of the
peaks is the same as for the dark matter field, at the peak location.
While this is a critical assumption in this paper, it is not
necessarily held for actual objects such as halos with relatively
small masses \cite{LP11}, in which case the velocities of tracers can
depend on the environment, as well as on the underlying smoothing
procedure in general. The inclusion of such dependences is beyond the
scope of this paper, and we assume the velocities of biased objects
are well approximated by those of peaks.

Using Eqs.~(\ref{eq:2-1-02})--(\ref{eq:2-1-04}), statistics of
momentum field (and cross correlations to the density field as well)
can be calculated in the way of iPT \cite{Mat11}. The full formulation
of extending the iPT including the momentum field is developed in
Sec.~\ref{sec:General}. In the rest of this section, we focus on its
implication to the Lagrangian velocity bias.

\subsection{Velocity bias in Lagrangian space}

The Fourier transform of Eq.~(\ref{eq:2-1-02}) is given by
\begin{equation}
  \bm{j}_X^\mathrm{L}(\bm{k})
  = \bm{v}^\mathrm{L}(\bm{k}) 
  + \int_{\bm{k}_1+\bm{k}_2=\bm{k}}
  \delta_X^\mathrm{L}(\bm{k}_1)\,
  \bm{v}^\mathrm{L}(\bm{k}_2),
  \label{eq:2-2-01}
\end{equation}
where we use a notation
\begin{equation}
  \int_{\bm{k}_1+\cdots +\bm{k}_2=\bm{k}} \cdots \equiv 
  \int \frac{d^3k_1}{(2\pi)^3}\cdots \frac{d^3k_n}{(2\pi)^3}\,
  \delta_\mathrm{D}^3(\bm{k}_1+\cdots +\bm{k}_n-\bm{k})\,
  \cdots.
  \label{eq:2-2-02}
\end{equation}
Taking a cross correlation with linear density field
$\delta_\mathrm{L}(\bm{k})$, we have
\begin{multline}
  \left\langle
    \delta_\mathrm{L}(\bm{k})
    \bm{j}_X^\mathrm{L}(\bm{k}')
  \right\rangle
  = \left\langle
    \delta_\mathrm{L}(\bm{k})
    \bm{v}^\mathrm{L}(\bm{k}') 
  \right\rangle
\\
    + \int_{\bm{k}_1+\bm{k}_2=\bm{k}'}
    \left\langle
        \delta_\mathrm{L}(\bm{k})
        \delta_X^\mathrm{L}(\bm{k}_1)\,
        \bm{v}^\mathrm{L}(\bm{k}_2)
        \right\rangle.
  \label{eq:2-2-03}
\end{multline}
Although the second term on the right-hand side (RHS) corresponds to a
higher-order correction in the usual sense of perturbation theory, what
we see below is that this term gives a contribution which linearly
biases the velocities of halos in the peaks model.

Equation~(\ref{eq:2-2-03}) is to be evaluated by means of iPT. For the
purpose that the results can be compared with those in the literature,
we consider a model where the displacement field is smoothed by a
filtering scale $R$, in accordance with most of the literature in
which smoothed underlying density and velocity fields are considered.
In this case, the perturbative expansion of the displacement field is
given by
\begin{equation}
  \bm{\varPsi}(\bm{k})
  = \sum_{n=1}^\infty
  \frac{i}{n!} \int_{\bm{k}_1+\cdots+\bm{k}_n=\bm{k}}
  \bar{\bm{L}}_n(\bm{k}_1,\ldots,\bm{k}_n)\,
  \delta_\mathrm{L}(\bm{k}_1) \cdots \delta_\mathrm{L}(\bm{k}_n),
  \label{eq:2-2-04}
\end{equation}
where $\bar{\bm{L}}_n$ is the smoothed, symmetrized kernel of
$n$th-order Lagrangian perturbation theory (LPT).

There are at least two schemes of smoothing the LPT kernels. The first
one is to smooth the linear density contrast $\delta_\mathrm{L}$ in
Lagrangian space, and the second one is to smooth the nonlinear
displacement field $\bm{\varPsi}$. Accordingly, the relation between
the LPT kernel $\bm{L}_n$ and the smoothed kernel $\bar{\bm{L}}_n$ is
given by either
\begin{equation}
  \bar{\bm{L}}_n(\bm{k}_1,\ldots,\bm{k}_n) =
  W(k_1R)\cdots W(k_nR)
  \bm{L}_n(\bm{k}_1,\ldots,\bm{k}_n),
  \label{eq:2-2-04a}
\end{equation}
or
\begin{equation}
  \bar{\bm{L}}_n(\bm{k}_1,\ldots,\bm{k}_n) =
  W(|\bm{k}_1+\cdots +\bm{k}_n|R)
  \bm{L}_n(\bm{k}_1,\ldots,\bm{k}_n).
  \label{eq:2-2-04b}
\end{equation}
The first-order kernel does not have any difference between the two
smoothing schemes above, and is given by
\begin{equation}
  \bar{\bm{L}}_1(\bm{k}) = W(kR)\bm{L}_1(\bm{k}).
  \label{eq:2-2-04c}
\end{equation}
In the following, we assume this relation for the first-order kernel,
and we leave the expression $\bar{\bm{L}}_n$ for higher-order kernels.
The LPT kernels up to third order are given by \cite{Cat95,CT96,Mat15}
\begin{align}
& \bm{L}^{(1)}(\bm{k}) = \frac{\bm{k}}{k^2},
\label{eq:2-2-04-1a}\\
& \bm{L}^{(2)}(\bm{k}_1,\bm{k}_2)
  =\frac37 \frac{\bm{k}_{12}}{{k_{12}}^2}
  \left[1 - \left(\frac{\bm{k}_1 \cdot \bm{k}_2}{k_1 k_2}\right)^2\right],
\label{eq:2-2-04-1b}\\
&  \bm{L}^{(3)}(\bm{k}_1,\bm{k}_2,\bm{k}_3) =
  \frac13
  \left[\bm{L}^{(3{\rm a})}(\bm{k}_1,\bm{k}_2,\bm{k}_3) + {\rm perm.}\right];
  \label{eq:2-2-04-1c}
\end{align}
where
\begin{align}
  & \bm{L}^{(3{\rm a})}(\bm{k}_1,\bm{k}_2,\bm{k}_3)
\nonumber\\
& \quad
  = \frac{\bm{k}_{123}}{{k_{123}}^2}
  \left\{
      \frac57
      \left[1 - \left(\frac{\bm{k}_1 \cdot \bm{k}_2}{k_1 k_2}\right)^2\right]
      \left[1 - \left(\frac{\bm{k}_{12} \cdot \bm{k}_3}
          {{k}_{12} k_3}\right)^2\right]
  \right.
\nonumber\\
& \qquad\quad
  \left.
  - \frac{1}{3}
  \left[
      1 - 3\left(\frac{\bm{k}_1 \cdot \bm{k}_2}{k_1 k_2}\right)^2
      +\, 2 \frac{(\bm{k}_1 \cdot \bm{k}_2)(\bm{k}_2 \cdot \bm{k}_3)
        (\bm{k}_3 \cdot \bm{k}_1)}{{k_1}^2 {k_2}^2 {k_3}^2}
  \right]\right\}
\nonumber\\
  & \qquad
    + \frac{3}{7}
    \frac{\bm{k}_{123}\times(\bm{k}_1\times\bm{k}_{23})}
    {{k_{123}}^2{k_1}^2{k_{23}}^2}
    (\bm{k}_1\cdot\bm{k}_{23})
    \left[
      1 - \left(\frac{\bm{k}_2\cdot\bm{k}_3}{k_2 k_3}\right)
    \right].
\label{eq:2-2-04-2}
\end{align}

Perturbative expansions of other variables are given by
\begin{align}
  \delta_X^\mathrm{L}(\bm{k})
  &=
  \sum_{n=1}^\infty
 \frac{1}{n!} \int_{\bm{k}_1+\cdots+\bm{k}_n=\bm{k}}
  b_X^{\mathrm{L}(n)}(\bm{k}_1,\ldots,\bm{k}_n)\,
  \delta_\mathrm{L}(\bm{k}_1) \cdots \delta_\mathrm{L}(\bm{k}_n),
  \label{eq:2-2-05a}
\\
  \bm{v}^\mathrm{L}(\bm{k})
  &=
  \sum_{n=1}^\infty
    \frac{iaHf}{(n-1)!}
    \nonumber \\
  &\quad\times
  \int_{\bm{k}_1+\cdots+\bm{k}_n=\bm{k}}
  \bar{\bm{L}}_n(\bm{k}_1,\ldots,\bm{k}_n)\,
  \delta_\mathrm{L}(\bm{k}_1) \cdots \delta_\mathrm{L}(\bm{k}_n),
  \label{eq:2-2-05b}
\end{align}
where $Hf = \dot{D}/D$, and $D$ is the linear growth factor. We have
used the approximate time dependence of the kernel function,
$\bm{L}_n \propto D^n$ \cite{Mat15}, and an identity
$d(D^n)/dt = n Hf D^n$. Substituting the expansion of
Eqs.~(\ref{eq:2-2-05a}) and (\ref{eq:2-2-05b}) into
Eq.~(\ref{eq:2-2-03}), and assuming Gaussian initial conditions, we
have
\begin{align}
  &
    \left\langle
    \delta_\mathrm{L}(\bm{k}_1)
    \bm{j}_X^\mathrm{L}(\bm{k})
  \right\rangle
  = \left\langle
    \delta_\mathrm{L}(\bm{k}_1)
    \bm{v}^{\mathrm{L}(1)}(\bm{k}) 
  \right\rangle
  + \left\langle
    \delta_\mathrm{L}(\bm{k}_1)
    \bm{v}^{\mathrm{L}(3)}(\bm{k}) 
  \right\rangle + \cdots
\nonumber \\  
  & \hspace{4pc} + \int_{\bm{k}'+\bm{k}''=\bm{k}}
  \left[
  \left\langle
    \delta_\mathrm{L}(\bm{k}_1)
    \delta_X^{\mathrm{L}(2)}(\bm{k}')\,
    \bm{v}^{\mathrm{L}(1)}(\bm{k}'')
    \right\rangle
    \right.
    \nonumber \\
  & \hspace{8pc} \left.
 +  \left\langle
    \delta_\mathrm{L}(\bm{k}_1)
    \delta_X^{\mathrm{L}(1)}(\bm{k}')\,
    \bm{v}^{\mathrm{L}(2)}(\bm{k}'')
  \right\rangle + \cdots
  \right]
\nonumber\\
  & \quad
    =  iaHf
    (2\pi)^3\delta_\mathrm{D}^3(\bm{k}_1+\bm{k})P_\mathrm{L}(k_1)
    \nonumber\\
  &\hspace{3pc} \times
  \Biggl[
    \bar{\bm{L}}_1(\bm{k}) +
    \frac{3}{2}
    \int \frac{d^3p}{(2\pi)^3}
    \bar{\bm{L}}_3(\bm{k},\bm{p},-\bm{p})
    P_\mathrm{L}(p)
    +\cdots
\nonumber\\
  & \hspace{4pc} +
  \int \frac{d^3p}{(2\pi)^3}
  c_X^{(2)}(\bm{k},\bm{p})
  \bar{\bm{L}}_1(-\bm{p})P_\mathrm{L}(p)
\nonumber\\
  & \hspace{4pc}
  + 2 \int \frac{d^3p}{(2\pi)^3}
  c_X^{(1)}(p)
  \bar{\bm{L}}_2(\bm{k},\bm{p})P_\mathrm{L}(p)
  + \cdots \Biggr],
  \label{eq:2-2-06}
\end{align}
where the higher-order self-loops of the bias functions are
renormalized according to the iPT. The functions $c_X^{(n)}$ are the
renormalized bias functions given by \cite{Mat11}
\begin{align}
  &
    c^{(n)}_X(\bm{k}_1,\ldots,\bm{k}_n)
    \nonumber \\
  & \equiv
  \sum_{m=0}^\infty
  \frac{1}{m!}
  \int \frac{d^3p_1}{(2\pi)^3} \cdots \frac{d^3p_m}{(2\pi)^3}
    b^{\mathrm{L}(n+m)}_X(\bm{k}_1,\ldots,\bm{k}_n,\bm{p}_1,\ldots,\bm{p}_m)
    \nonumber\\
  & \hspace{11pc} \times
  \left\langle
    \delta_\mathrm{L}(\bm{p}_1) \cdots \delta_\mathrm{L}(\bm{p}_m)
  \right\rangle
  \nonumber\\
  &=
  \sum_{m=0}^\infty
  \frac{1}{2^mm!}
  \int \frac{d^3p_1}{(2\pi)^3} \cdots \frac{d^3p_m}{(2\pi)^3}
    \nonumber\\
  & \hspace{4pc} \times
  b^{\mathrm{L}(n+2m)}_X(\bm{k}_1,\ldots,\bm{k}_n,
  \bm{p}_1,-\bm{p}_1,\ldots,\bm{p}_m,-\bm{p}_m)
    \nonumber\\
  & \hspace{4pc} \times
  P_\mathrm{L}(p_1) \cdots P_\mathrm{L}(p_m).
  \label{eq:2-2-06-1a}
\end{align}
The first line is derived from the general definition
[Eq.~(\ref{eq:3-2-01}) below], and the second line is derived with
Gaussian initial conditions. For example,
\begin{align}
  &
  c^{(1)}_X(k) = 
  b^{\mathrm{L}(1)}_X(k)
  + \frac{1}{2} \int \frac{d^3p}{(2\pi)^3}
  b^{\mathrm{L}(3)}_X(\bm{k},\bm{p},-\bm{p})
    P_\mathrm{L}(p)
    \nonumber\\
  & \hspace{3pc}
  + \frac{1}{8} \int \frac{d^3p}{(2\pi)^3} \frac{d^3p'}{(2\pi)^3}
    b^{\mathrm{L}(5)}_X(\bm{k},\bm{p},-\bm{p},\bm{p}',-\bm{p}')
    \nonumber\\
  & \hspace{10pc} \times
  P_\mathrm{L}(p) P_\mathrm{L}(p') + \cdots,
  \label{eq:2-2-06-2a}\\
  & c^{(2)}_X(\bm{k}_1,\bm{k}_2)
  = b^{\mathrm{L}(2)}_X(\bm{k}_1,\bm{k}_2)
    \nonumber\\
  & \hspace{3pc}
  + \frac{1}{2} \int \frac{d^3p}{(2\pi)^3}
  b^{\mathrm{L}(4)}_X(\bm{k}_1,\bm{k}_2,\bm{p},-\bm{p})
  P_\mathrm{L}(p)
  + \cdots,
  \label{eq:2-2-06-2b}
\end{align}
and so forth. We have used the fact that the self-contractions vanish
$\langle\bm{v}^{\mathrm{L}(n)}\rangle = \langle
\delta_X^{\mathrm{L}(n)} \rangle = 0$ for every order. The
renormalized bias functions of the peaks model up to the second order
are given by \cite{Des13,LMD15,MD16}
\begin{align}
  \label{eq:2-2-07a}
  & c_X^{(1)}(k) = 
  \left( b_{10} + b_{01} k^2 \right) W(kR),
\\
  \label{eq:2-2-07b}
  & c_X^{(2)}(\bm{k}_1, \bm{k}_2) =
  \left\{
    b_{20} + b_{11}\left({k_1}^2 + {k_2}^2\right) + b_{02} {k_1}^2 {k_2}^2
    \right.
    \nonumber \\
  & \hspace{5pc}
    \left.
  - 2 \chi_1 (\bm{k}_1\cdot\bm{k}_2)
    +\, \omega_{10}\left[3(\bm{k}_1\cdot\bm{k}_2)^2 - {k_1}^2{k_2}^2\right]
  \right\}
    \nonumber\\
  & \hspace{11pc}
  \times W(k_1R)W(k_2R),
\end{align}
where coefficients $b_{ij}$, $\chi_1$ and $\omega_{10}$ are
scale-independent constants which depend on the threshold of the peaks
(see Refs.~\cite{LMD15,MD16} for their definitions).

The second and fourth terms in the square bracket of
Eq.~(\ref{eq:2-2-06}) are the usual mode-coupling terms. In the third
term, however, we have $\bar{\bm{L}}_1(\bm{p}) = W(pR)\bm{p}/p^2$, and
the integration over the angle of $\bm{p}$ leaves the term with only
$\chi_1$ of Eq.~(\ref{eq:2-2-07b}), resulting in
\begin{align}
  &
 \int \frac{d^3p}{(2\pi)^3} c_X^{(2)}(\bm{k},\bm{p})
 \bar{\bm{L}}_1(-\bm{p})P_\mathrm{L}(p)
    \nonumber\\
  &\qquad
 = 2 \chi_1 W(kR)
  \int \frac{d^3p}{(2\pi)^3}
  \frac{(\bm{k}\cdot\bm{p})\bm{p}}{p^2} P_\mathrm{L}(p)W^2(pR)
    \nonumber\\
  &\qquad
  = \frac{2}{3} \chi_1 \bm{k} W(kR)
  \int \frac{p^2dp}{2\pi^2} P_\mathrm{L}(p) W^2(pR)
    \nonumber\\
  &\qquad
  = \frac{2}{3} \chi_1 {\sigma_0}^2 \bm{k} W(kR)
  = -{R_\mathrm{v}}^2 \bm{k} W(kR)
    \nonumber\\
  &\qquad
  = -{R_\mathrm{v}}^2 k^2 \bar{\bm{L}}_1(\bm{k}),
  \label{eq:2-2-08}
\end{align}
where we have used
$(4\pi)^{-1}\int d\varOmega_p p_ip_j/p^2 = \delta_{ij}/3$ and
$\chi_1 = -3/(2{\sigma_1}^2)$,
${R_\mathrm{v}}^2 = {\sigma_0}^2/{\sigma_1}^2$, and spectral
parameters $\sigma_n(R)$ are defined by
\begin{equation}
  \label{eq:2-2-08-1}
  {\sigma_n}^2 \equiv
  \int \frac{k^2dk}{2\pi^2} k^{2n} P_\mathrm{L}(k) W^2(kR).
\end{equation}
The above Eq.~(\ref{eq:2-2-08}) has the same functional form as the
linear term. Thus, Eq.~(\ref{eq:2-2-06}) reduces to
\begin{multline}
  \left\langle
    \delta_\mathrm{L}(\bm{k}_1)
    \bm{j}_X^\mathrm{L}(\bm{k})
  \right\rangle
  =  iaHf (2\pi)^3\delta_\mathrm{D}^3(\bm{k}_1+\bm{k})P_\mathrm{L}(k)
  \\ \times
  \left[
    \left(1 - {R_\mathrm{v}}^2k^2\right) \bar{\bm{L}}_1(\bm{k}) + \mbox{m.c.}
  \right],
  \label{eq:2-2-09}
\end{multline}
where ``$+\,\mbox{m.c.}$'' represents mode-coupling terms. The
correlator for the matter momentum field
$\langle\delta_\mathrm{L}\bm{j}_\mathrm{m}^\mathrm{L}\rangle$ is just
given by putting ${R_\mathrm{v}}^2=0$ and $W(kR)=1$ in the above
expression. Thus we have
\begin{equation}
  \left\langle
    \delta_\mathrm{L}(\bm{k}_1)
    \bm{j}_X^\mathrm{L}(\bm{k})
  \right\rangle = 
  \left(1 - R_\mathrm{v}^2 k^2\right) W(kR)
  \left\langle
    \delta_\mathrm{L}(\bm{k}_1)
    \bm{j}_\mathrm{m}^\mathrm{L}(\bm{k})
  \right\rangle + \mbox{m.c.}
  \label{eq:2-2-10}
\end{equation}
In this way, the $\chi_1$ term in $c_X^{(2)}$ of the peaks model
introduces a linear velocity bias. The scale-dependent velocity bias
factor, $(1-{R_\mathrm{v}}^2k^2) W(kR)$, is consistent with the
finding of Ref.~\cite{Des08}. Equation~(\ref{eq:2-2-10}) corresponds
to Eqs.~(2) and (9) of Ref.~\cite{Bal15}. In the following, we use the
notation,
\begin{equation}
  b_\mathrm{v}(k) \equiv 
  \left( 1 - {R_\mathrm{v}}^2k^2\right) W(kR)
  \label{eq:2-2-11}
\end{equation}
which represents the lowest-order effect of velocity bias in Lagrangian
space.

\subsection{Relation to the one-loop density power spectra of iPT}

The same type of the seemingly linear term is also contained in a
one-loop correction of iPT power spectrum. In the iPT, the normalized
two-point propagator is given by \cite{Mat14}
\begin{multline}
  \hat{\varGamma}_X^{(1)}(\bm{k})
  = c_X^{(1)}(k) + \bm{k}\cdot\bar{\bm{L}}_1(\bm{k})
  \\
  + \int\frac{d^3p}{(2\pi)^3} P_{\rm L}(p)
  \biggl\{
      c_X^{(2)}(\bm{k},\bm{p})\,
      \bm{k}\cdot\bar{\bm{L}}_1(-\bm{p})
      \\
      +
      \left[
        c_X^{(1)}(p)
        + \bm{k}\cdot\bar{\bm{L}}_1(\bm{p})
      \right]
      \bm{k}\cdot\bar{\bm{L}}_2(\bm{k},-\bm{p})
      \\
      + \frac12
      \bm{k}\cdot\bar{\bm{L}}_3(\bm{k},\bm{p},-\bm{p})
  \biggr\},
\label{eq:2-3-1}
\end{multline}
where the displacement field is filtered by a smoothing scale $R$ in
accordance with the model of the previous subsection, and we drop a term
which vanishes for an isotropic function $c^{(1)}_X(p)$. The second
and third terms in the integrand are mode-coupling terms. The first
term has essentially the same form as the third term of
Eq.~(\ref{eq:2-2-06}), which is the very source of the velocity bias.
After the integration, this term is equal to
$-{R_\mathrm{v}}^2k^2\bm{k}\cdot\bar{\bm{L}}_1(\bm{k})$. Hence, the
first term in the curly bracket of Eq.~(\ref{eq:2-3-1}) gives a term
without mode coupling, and we have
\begin{equation}
    \hat{\varGamma}_X^{(1)}(\bm{k}) = 
    \left(1 - {R_\mathrm{v}}^2 k^2\right)
    \bm{k}\cdot\bar{\bm{L}}_1(\bm{k})
    + c_X^{(1)}(k)
    + \mbox{m.c.}
  \label{eq:2-3-2}
\end{equation}

In real space without the redshift-space distortions, we have
$\bm{L}_1(\bm{k}) = \bm{k}/k^2$. Consequently, the Eulerian linear
bias factor is effectively given by
\begin{equation}
    b(k) = b_\mathrm{v}(k) +  c_X^{(1)}(k)
    = 
    \left[
        1 + b_{10} + \left(b_{01} - {R_\mathrm{v}}^2\right) k^2
    \right] W(kR).
  \label{eq:2-3-3}
\end{equation}
Therefore, the $\chi_1$ term again modifies the linear bias, not only
for the velocity field, but also for the density field.

In Eq.~(\ref{eq:2-3-3}), the magnitude of the scale-dependent part
$\propto k^2$ of the linear bias factor is reduced compared with the
prediction of the linear theory. This effect can be interpreted from
the presence of velocity bias, because the velocities of halos are
slower than those of dark matter on average, and halos are less
clustered than in the case without velocity bias. More quantitatively,
one can see this as follows: consider the simplest case when the halos
are not linearly biased, $b_{10}=b_{01}=0$, and formations and
destructions of halos can be neglected at the lowest order. In this
ideal case, the number density of halos are approximately conserved at
the lowest order,
$\dot{\delta}_X + a^{-1}i\bm{k}\cdot\bm{j}_X \sim 0$. Because of the
velocity bias, we have
$\bm{k}\cdot\bm{j}_X \sim (1-{R_\mathrm{v}}^2k^2)W(kR)
\bm{k}\cdot\bm{j}_\mathrm{m}$. Combining these relations with the mass
conservation equation
$\dot{\delta}_\mathrm{m} + a^{-1}i\bm{k}\cdot\bm{j}_\mathrm{m}=0$, we
have a relation,
$\dot{\delta}_X \sim
(1-{R_\mathrm{v}}^2k^2)W(kR)\dot{\delta}_\mathrm{m}$, which indicates
$b(k) \sim (1-{R_\mathrm{v}}^2k^2)W(kR)$, and this is consistent to
Eq.~(\ref{eq:2-3-3}) within our assumption $b_{10}=b_{01}=0$. In this
way, the amplitude of the scale-dependent part of the linear bias
factor is reduced due to the velocity bias. All such effects are
already included in the iPT with the peak constraint.

\subsection{Propagators in Lagrangian space}

The correlator of Eq.~(\ref{eq:2-2-06}) is related to the two-point
propagators \cite{BCS08}. We define the two-point propagator of the
momentum field in Lagrangian space,
$\bm{\varGamma}^{\mathrm{vL}(1)}_X(\bm{k})$, by
\begin{equation}
  \left\langle
    \frac{\delta\bm{j}_X^\mathrm{L}(\bm{k})}
    {\delta\delta_\mathrm{L}(\bm{k}_1)} 
  \right\rangle =
  i\,(2\pi)^3\delta_\mathrm{D}^3(\bm{k}-\bm{k}_1)
  \bm{\varGamma}^{\mathrm{vL}(1)}_X(\bm{k}_1),
  \label{eq:2-4-01}
\end{equation}
where $\delta/\delta\delta_\mathrm{L}(\bm{k})$ is the functional
derivative with respect to the linear density field
$\delta_\mathrm{L}(\bm{k})$ in Fourier space, and the appearance of
the Dirac's delta function is a consequence of translation
invariance of space. The imaginary unit $i$ is put in the definition,
so as to make the propagator real vector. In Gaussian initial
conditions, we have a relation,
\begin{align}
  \left\langle
    \delta_\mathrm{L}(\bm{k}_1)
    \bm{j}_X^\mathrm{L}(\bm{k})
  \right\rangle_\mathrm{c}
  &=
  P_\mathrm{L}(k_1)
  \left\langle
    \frac{\delta\bm{j}_X^\mathrm{L}(\bm{k})}
    {\delta\delta_\mathrm{L}(-\bm{k}_1)} 
    \right\rangle
    \nonumber\\
    &=
  -i\,(2\pi)^3\delta_\mathrm{D}^3(\bm{k}+\bm{k}_1)
  P_\mathrm{L}(k_1)
  \bm{\varGamma}^{\mathrm{vL}(1)}_X(\bm{k}_1),
  \label{eq:2-4-02}
\end{align}
where $\langle\cdots\rangle_\mathrm{c}$ represents the cumulants of
the second order, and we have
$\bm{\varGamma}^{\mathrm{vL}(1)}_X(-\bm{k})=-\bm{\varGamma}^{\mathrm{vL}(1)}_X(\bm{k})$
from the parity symmetry.

Including the mode-coupling terms of Eqs.~(\ref{eq:2-2-06}) and
(\ref{eq:2-2-09}), the two-point correlators of the momentum field in
Lagrangian space is given by
\begin{multline}
  \bm{\varGamma}^{\mathrm{vL}(1)}_X(\bm{k})
  = aHf
  \Biggl\{
  \left(1 - {R_\mathrm{v}}^2k^2\right) \bar{\bm{L}}_1(\bm{k})
  \\
    + \frac{1}{2} \int \frac{d^3p}{(2\pi)^3} P_\mathrm{L}(p)
    \left[
      4 c^{(1)}_X(p)
      \bar{\bm{L}}_2(\bm{k},\bm{p})
      + 3 \bar{\bm{L}}_3(\bm{k},\bm{p},-\bm{p})
    \right]
  \Biggr\}.
  \label{eq:2-4-04-1}
\end{multline}

Similarly, we consider the three-point propagator in Lagrangian space,
$\bm{\varGamma}^{\mathrm{vL}(2)}_X(\bm{k}_1,\bm{k}_2)$, which is
defined by
\begin{equation}
  \left\langle
    \frac{\delta^2\bm{j}_X^\mathrm{L}(\bm{k})}
    {\delta\delta_\mathrm{L}(\bm{k}_1)\delta\delta_\mathrm{L}(\bm{k}_2)} 
  \right\rangle = i\,
  (2\pi)^3\delta_\mathrm{D}^3(\bm{k}-\bm{k}_1-\bm{k}_2)
  \bm{\varGamma}^{\mathrm{vL}(2)}_X(\bm{k}_1,\bm{k}_2).
  \label{eq:2-4-05}
\end{equation}
In Gaussian initial conditions, we have a relation,
\begin{align}
&  \left\langle
    \delta_\mathrm{L}(\bm{k}_1)
    \delta_\mathrm{L}(\bm{k}_2)
    \bm{j}_X^\mathrm{L}(\bm{k})
  \right\rangle_\mathrm{c}
  \nonumber \\
  & \qquad =
  P_\mathrm{L}(k_1) P_\mathrm{L}(k_2)
  \left\langle
    \frac{\delta^2\bm{j}_X^\mathrm{L}(\bm{k})}
    {\delta\delta_\mathrm{L}(-\bm{k}_1)\delta\delta_\mathrm{L}(-\bm{k}_2)} 
  \right\rangle
  \nonumber \\
  & \qquad = -i\,
  (2\pi)^3\delta_\mathrm{D}^3(\bm{k}+\bm{k}_1+\bm{k}_2)
  P_\mathrm{L}(k_1) P_\mathrm{L}(k_2)
  \bm{\varGamma}^{\mathrm{vL}(2)}_X(\bm{k}_1,\bm{k}_2).
  \label{eq:2-4-06}
\end{align}
Thus, it is sufficient to evaluate the left-hand side (LHS) of the
above equation to obtain an expression of the three-point propagator.

From Eq.~(\ref{eq:2-2-01}), we have
\begin{multline}
  \left\langle
    \delta_\mathrm{L}(\bm{k}_1)
    \delta_\mathrm{L}(\bm{k}_2)
    \bm{j}_X^\mathrm{L}(\bm{k})
  \right\rangle_\mathrm{c}
  = \left\langle
    \delta_\mathrm{L}(\bm{k}_1)
    \delta_\mathrm{L}(\bm{k}_2)
    \bm{v}^\mathrm{L}(\bm{k}) 
  \right\rangle_\mathrm{c}
 \\
    + \int_{\bm{k}'+\bm{k}''=\bm{k}}
    \left\langle
        \delta_\mathrm{L}(\bm{k}_1)
        \delta_\mathrm{L}(\bm{k}_2)
        \left[
           \delta_X^\mathrm{L}(\bm{k}')\,
           \bm{v}^\mathrm{L}(\bm{k}'')
       \right]
        \right\rangle_\mathrm{c}.
  \label{eq:2-4-07}
\end{multline}
The cumulants in the above equation are all third order, where the
square bracket in the cumulant of the last term is considered as a
single variable. Substituting the expansion of Eqs.~(\ref{eq:2-2-05a})
and (\ref{eq:2-2-05b}) into Eq.~(\ref{eq:2-4-07}), we have
\begin{align}
&  \left\langle
    \delta_\mathrm{L}(\bm{k}_1)
    \delta_\mathrm{L}(\bm{k}_2)
    \bm{j}_X^\mathrm{L}(\bm{k})
  \right\rangle_\mathrm{c}
  = \left\langle
    \delta_\mathrm{L}(\bm{k}_1)
    \delta_\mathrm{L}(\bm{k}_2)
    \bm{v}^{\mathrm{L}(2)}(\bm{k}) 
  \right\rangle_\mathrm{c} + \cdots
\nonumber\\
& \qquad
  + \int_{\bm{k}'+\bm{k}''=\bm{k}}
  \left\{
    \left\langle
      \delta_\mathrm{L}(\bm{k}_1)
      \delta_\mathrm{L}(\bm{k}_2)
      \left[
        \delta_X^{\mathrm{L}(1)}(\bm{k}')\,
        \bm{v}^{\mathrm{L}(1)}(\bm{k}'')
              \right]\right\rangle_\mathrm{c}
              \right.
\nonumber\\
  & \hspace{5pc}
    \left.
    +
    \left\langle
      \delta_\mathrm{L}(\bm{k}_1)
      \delta_\mathrm{L}(\bm{k}_2)
      \left[
        \delta_X^{\mathrm{L}(3)}(\bm{k}')\,
        \bm{v}^{\mathrm{L}(1)}(\bm{k}'')
      \right]
    \right\rangle_\mathrm{c} + \cdots
  \right\}
\nonumber\\
  & \quad
    = -iaHf (2\pi)^3
  \delta_\mathrm{D}^3(\bm{k}+\bm{k}_1+\bm{k}_2)
  P_\mathrm{L}(k_1)P_\mathrm{L}(k_2)
\nonumber\\
  &\qquad
  \times
  \Biggl[
    2\bar{\bm{L}}_2(\bm{k}_1,\bm{k}_2)
    + c^{(1)}_X(k_1) \bar{\bm{L}}_1(\bm{k}_2)
    + c^{(1)}_X(k_2) \bar{\bm{L}}_1(\bm{k}_1)
\nonumber\\
  &\hspace{3pc}
    + \int\frac{d^3p}{(2\pi)^3}
    c^{(3)}_X(\bm{k}_1,\bm{k}_2,\bm{p})
    \bar{\bm{L}}_1(-\bm{p})P_\mathrm{L}(p)
    + \cdots
  \Biggr],
  \label{eq:2-4-08}
\end{align}
where the higher-order self-loops of the bias functions are
renormalized according to the iPT.

The fourth term in the square bracket is similar to the integral of
Eq.~(\ref{eq:2-2-08}). The third-order renormalized bias functions of
the peaks model is given in Ref.~\cite{LMD15}. The terms of odd parity
with respect to the third argument in their expression are given
by\footnote{In the $c_{00001}$ term of Eq.~(91) of Ref.~\cite{LMD15},
  the term $(\bm{k}_1\cdot\bm{k}_2)k_3^2$ should be replaced by
  $(\bm{k}_1\cdot\bm{k}_2)^2k_3^2$, i.e., this term has even parity.}
\begin{multline}
  c^{(3)}_X(\bm{k}_1,\bm{k}_2,\bm{p})
  \supset 
  -2 W(k_1R)W(k_2R)W(pR)
  \\ \times
  \left[
    c_{10100}\,(\bm{k}_1+\bm{k}_2)\cdot\bm{p}
    + c_{01100}\,
    \left({k_2}^2\bm{k}_1 + {k_1}^2\bm{k}_2
    \right)\cdot\bm{p}
  \right].
  \label{eq:2-4-09}
\end{multline}
The coefficients can be calculated by their definition, and the
results are given by $c_{10100} = -3b_{10}/2{\sigma_1}^2$,
$c_{01100} = -3b_{01}/2{\sigma_1}^2$. Other terms with even parity do
not contribute to the integral. Performing the similar calculation of
Eq.~(\ref{eq:2-2-08}), we have
\begin{align}
&  \int\frac{d^3p}{(2\pi)^3}
  c^{(3)}_X(\bm{k}_1,\bm{k}_2,\bm{p})
  \bar{\bm{L}}_1(-\bm{p})P_\mathrm{L}(p)
  \nonumber \\
& \qquad  = -{R_\mathrm{v}}^2
  \left( b_{10} + b_{01}{k_1}^2 \right) \bm{k}_2 W(k_1R) W(k_2R)
  + (\bm{k}_1 \leftrightarrow \bm{k}_2)
  \nonumber \\
&\qquad  = -c^{(1)}_X(k_1)  {R_\mathrm{v}}^2 {k_2}^2 \bar{\bm{L}}_1(\bm{k}_2)
  + (\bm{k}_1 \leftrightarrow \bm{k}_2).
  \label{eq:2-4-10}
\end{align}
Consequently, the three-point propagator of Eq.~(\ref{eq:2-4-05}) is
given by
\begin{multline}
  \bm{\varGamma}^{\mathrm{vL}(2)}_X(\bm{k}_1,\bm{k}_2) =
  aHf
  \left[
    c^{(1)}_X(k_1)
    \left(
      1 - {R_\mathrm{v}}^2{k_2}^2
    \right)
    \bar{\bm{L}}_1(\bm{k}_2)
    \right.
      \\
    \left.
    + (\bm{k}_1 \leftrightarrow \bm{k}_2)
    +
    2\bar{\bm{L}}_2(\bm{k}_1,\bm{k}_2)
  \right],
  \label{eq:2-4-12}
\end{multline}
up to the lowest-order approximation.

\subsection{The displacement field}

For the purpose of comparison, we consider the statistics of the
displacement field. In the calculations above, the momentum field can
be replaced by a displacement field,
$a\dot{\bm{\varPsi}} \rightarrow \bm{\varPsi}$, and we can follow
almost the same steps. The number-weighted displacement field
$\bm{\psi}_X^\mathrm{L}$ in Lagrangian space is defined by
\begin{equation}
  \bm{\psi}_X^\mathrm{L}(\bm{q}) = [1 + \delta_X^\mathrm{L}(\bm{q})]
  \bm{\varPsi}(\bm{q}),
  \label{eq:2-5-01}
\end{equation}
and the perturbative expansions of the displacement field in
Lagrangian space is given by Eq.~(\ref{eq:2-2-04}). Comparing these
equations with Eqs.~(\ref{eq:2-1-02}), (\ref{eq:2-1-04}) and
(\ref{eq:2-2-05b}), we see that replacing the Lagrangian kernels
$\bar{\bm{L}}_n \rightarrow \bar{\bm{L}}_n/naHf$ in the expression of
the above results for $\bm{j}_X^\mathrm{L}$ can give the results
for $\bm{\psi}_X^\mathrm{L}$. Following the calculations to derive
Eqs.~(\ref{eq:2-4-04-1}) and (\ref{eq:2-4-12}), the two- and
three-point propagators of displacement field are derived as
\begin{align}
  & \bm{\varGamma}^{\mathrm{dL}(1)}_X(\bm{k})
  =
    \left(1 - {R_\mathrm{v}}^2k^2\right) \bar{\bm{L}}_1(\bm{k})
    \nonumber\\
  & \quad
  + \frac{1}{2} \int \frac{d^3p}{(2\pi)^3} P_\mathrm{L}(p)
  \left[
    2 c^{(1)}_X(p) \bar{\bm{L}}_2(\bm{k},\bm{p}) +
    \bar{\bm{L}}_3(\bm{k},\bm{p},-\bm{p})
  \right],
  \label{eq:2-5-04a}\\
&  \bm{\varGamma}^{\mathrm{dL}(2)}_X(\bm{k}_1,\bm{k}_2) =
  c^{(1)}_X(k_1)
  \left(
    1 - {R_\mathrm{v}}^2{k_2}^2
  \right)
  \bar{\bm{L}}_1(\bm{k}_2)
  \nonumber\\
 & \hspace{8pc}
   + (\bm{k}_1 \leftrightarrow \bm{k}_2)
  + \bar{\bm{L}}_2(\bm{k}_1,\bm{k}_2).
  \label{eq:2-5-04b}
\end{align}

\section{\label{sec:VelocityDispersion}
  The velocity dispersion of peaks
}

\subsection{Nonlinear corrections to the velocity dispersion of peaks}

As an example of applications of the velocity bias derived from the
framework of iPT in the previous subsection, we consider the velocity
dispersion of peaks in this section. The velocity dispersion of biased
objects is defined by
\begin{equation}
  \sigma_\mathrm{nv}^2 \equiv
  \frac{1}{N_\mathrm{obj}} 
  \sum_{a=1}^{N_\mathrm{obj}} \left|\bm{v}(\bm{x}_a)\right|^2
  =
  \frac{1}{N_\mathrm{obj}} 
  \sum_{a=1}^{N_\mathrm{obj}} \left|\bm{v}^\mathrm{L}(\bm{q}_a)\right|^2,
  \label{eq:2-6-01}
\end{equation}
where $\bm{v}(\bm{x}_a)$ is the Eulerian velocity at Eulerian
coordinates $\bm{x}_a$ of an object labeled by $a$, and
$N_\mathrm{obj}$ is the total number of biased objects in a sample. In
the last expression, $\bm{v}^\mathrm{L}(\bm{q}_a)$ is the Lagrangian
velocity at Lagrangian coordinates $\bm{q}_a$. The last
equality holds because the Lagrangian velocity field is defined by
$\bm{v}^\mathrm{L}(\bm{q}) \equiv \bm{v}(\bm{x})$, where
$\bm{x}=\bm{q} + \bm{\varPsi}(\bm{q})$. Therefore, there are not any
differences between velocity dispersions of objects in Eulerian space
and in Lagrangian space.

The velocity dispersion of objects is not identical to the dispersion
of velocity field $\langle |\bm{v}(\bm{x})|^2 \rangle$. Instead, it is
given by a number-weighted dispersion of velocity field,
$\langle n_X(\bm{x})|\bm{v}(\bm{x})|^2\rangle/\bar{n}_X$. The velocity
dispersion of Eq.~(\ref{eq:2-6-01}) is given by an expression,
\begin{equation}
  \sigma_\mathrm{nv}^2 =
  \left\langle
      \left[1 + \delta^\mathrm{L}_X(\bm{q})\right]
      \left|\bm{v}^\mathrm{L}(\bm{q})\right|^2
  \right\rangle.
  \label{eq:2-6-02}
\end{equation}
This expression can be evaluated by applying the perturbative
expansions of Eqs.~(\ref{eq:2-2-05a}) and (\ref{eq:2-2-05b}), and by
following the similar calculations to derive the correlators of
Eqs.~(\ref{eq:2-2-06}) and (\ref{eq:2-4-12}). Just as in that case,
only terms of odd parity in the higher-order bias functions
$c^{(2)}_X$ and $c^{(3)}_X$ survive and they are represented by
${R_\mathrm{v}}^2$ and $c^{(1)}_X$ because of Eqs.~(\ref{eq:2-2-08})
and (\ref{eq:2-4-10}). Alternatively, Eq.~(\ref{eq:2-6-02}) is
equivalent to an expression
$\sigma_\mathrm{nv}^2 = \langle
\bm{v}^\mathrm{L}\cdot\bm{j}^\mathrm{L}_X \rangle$, and one can
evaluate this expression by applying the perturbative expansion of
Eq.~(\ref{eq:2-2-05b}) and using the results of the correlators,
Eqs.~(\ref{eq:2-4-04-1}) and (\ref{eq:2-4-12}). In either way, one can
derive the same result,
\begin{multline}
    \frac{\sigma_\mathrm{nv}^2}{a^2H^2f^2} = 
    \sigma_\mathrm{dpk}^2
 +  \int\frac{d^3k}{(2\pi)^3} \frac{d^3p}{(2\pi)^3}
 P_\mathrm{L}(k) P_\mathrm{L}(p)
 \\ \times
    \Biggl\{
      2 \left| \bar{\bm{L}}_2(\bm{k},\bm{p}) \right|^2
      + \frac{W(kR)}{k^2}
      \left(1 - \frac{1}{2}{R_\mathrm{v}}^2k^2\right)
      \\
     \times
      \left[
        4 c^{(1)}_X(p)\,\bm{k}\cdot\bar{\bm{L}}_2(\bm{k},\bm{p}) +
        3 \bm{k}\cdot\bar{\bm{L}}_3(\bm{k},\bm{p},-\bm{p})
      \right]
    \Biggr\},
    \label{eq:2-6-03}
\end{multline}
where
\begin{align}
  \sigma_\mathrm{dpk}^2
  &\equiv 
  \int\frac{dk}{2\pi^2} P_\mathrm{L}(k) W^2(kR)
    \left(1 - {R_\mathrm{v}}^2k^2\right)
    \nonumber\\
  &= {\sigma_{-1}}^2 - {R_\mathrm{v}}^2{\sigma_0}^2
  = {\sigma_{-1}}^2 - \frac{{\sigma_0}^4}{{\sigma_1}^2}.
  \label{eq:2-6-04}
\end{align}
This parameter $\sigma_\mathrm{dpk}^2$ is the ``peak displacement
dispersion'' defined in Ref.~\cite{Bal15}.

\subsection{Reducing the dimensions of integral}

The six-dimensional integrals in Eq.~(\ref{eq:2-6-03}) can be
evaluated by a combination of one-dimensional integrals as shown
below. Recently it is pointed out that the multidimensional
integrations appeared in the perturbation theory of the nonlinear
power spectrum can be evaluated by combining only one-dimensional (1D)
integrations of Hankel transforms with FFT \cite{SVM16, SV16, MFHB16,
  FBMH16}. The essence of the method is the realization that angular
parts of the multidimensional integrations can be analytically
performed, and all the remaining integrations can be represented by a
set of 1D Hankel transforms. Essentially the same method can be
applied here.

The particular forms of LPT kernels in Eq.~(\ref{eq:2-6-03}) are given
by
\begin{align}
  \left|\bar{\bm{L}}_2(\bm{k},\bm{p})\right|^2
  &= 
  \frac{9}{49}
  \frac{1}{|\bm{k}+\bm{p}|^2}
  \left[
    1 - \left(\frac{\bm{k}\cdot\bm{p}}{kp}\right)^2
  \right]^2
   W^2(kR) W^2(pR),
  \label{eq:2-6-04a}\\
  \bm{k}\cdot\bar{\bm{L}}_2(\bm{k},\bm{p})
  &=
  \frac{3}{7}
  \frac{\bm{k}\cdot(\bm{k}+\bm{p})}{|\bm{k}+\bm{p}|^2}
  \left[
    1 - \left(\frac{\bm{k}\cdot\bm{p}}{kp}\right)^2
  \right] W(kR) W(pR),
  \label{eq:2-6-04b}\\
  \bm{k}\cdot\bar{\bm{L}}_3(\bm{k},\bm{p},-\bm{p})
  &= \frac{5}{21}
  \left(
    \frac{k^2}{|\bm{k}+\bm{p}|^2} + \frac{k^2}{|\bm{k}-\bm{p}|^2}
  \right)
   \nonumber\\
  & \quad
    \times
  \left[
    1 - \left(\frac{\bm{k}\cdot\bm{p}}{kp}\right)^2
  \right]^2
   W(kR) W^2(pR),
  \label{eq:2-6-04c}
\end{align}
where we assume the smoothing scheme of linear density field,
Eq.~(\ref{eq:2-2-04a}). When the smoothing scheme of displacement,
Eq.~(\ref{eq:2-2-04b}), is adopted, the products of smoothing kernels
in Eqs.~(\ref{eq:2-6-04a})--(\ref{eq:2-6-04c}) are, respectively,
replaced by $W^2(kR)W^2(pR) \rightarrow W^2(|\bm{k}+\bm{p}|R)$,
$W(kR)W(pR) \rightarrow W(|\bm{k}+\bm{p}|R)$, and
$W(kR)W^2(pR) \rightarrow W(kR)$.

The angular integration of the variable $\bm{p}$ can be evaluated by a
following formula for an arbitrary function $F(k)$ and Legendre
polynomials $\mathsf{P}_l(x)$:
\begin{multline}
  \int \frac{d\varOmega_{\bm{p}}}{4\pi}
  F\left(|\bm{k}-\bm{p}|\right)
  \mathsf{P}_l\left(\frac{\bm{k}\cdot\bm{p}}{kp}\right)
  \\
  = 4\pi
  \int_0^\infty r^2 dr j_l(kr) j_l(pr)
  \int_0^\infty \frac{k'^2dk'}{2\pi^2} j_0(k'r) F(k'),
  \label{eq:2-6-05}
\end{multline}
which can be shown, e.g., by applying a 3D Fourier transform of
$F(k)$, the plane wave expansion of $e^{i\bm{k}\cdot\bm{r}}$, the
addition formula of Legendre polynomials, $\mathsf{P}_l(x)$, and the
orthogonality relation of spherical harmonics, $Y_l^m(\varOmega)$. The
last integral is a Hankel transform of $F(k)$. For example,
for $F(k)=1/k^2$, we have
\begin{equation}
  \int_0^\infty \frac{k^2dk}{2\pi^2} j_0(kr) \frac{1}{k^2}
  = \frac{1}{4\pi r}.
  \label{eq:2-6-06}
\end{equation}
Equations~(\ref{eq:2-6-05}) and (\ref{eq:2-6-06}) are sufficient to
evaluate the angular integrations of
Eqs.~(\ref{eq:2-6-04a})--(\ref{eq:2-6-04c}), by changing the
integration variable as $\bm{p} \rightarrow -\bm{p}$ when necessary.
The angular variable $\mu =\bm{k}\cdot\bm{p}/kp$ in those equations
are represented by Legendre polynomials as
$(1-\mu^2)^2 = (8/15)\mathsf{P}_0(\mu) - (16/21)\mathsf{P}_2(\mu) +
(8/35)\mathsf{P}_4(\mu)$,
$1-\mu^2 = (2/3)\mathsf{P}_0(\mu) - (2/3)\mathsf{P}_2(\mu)$ and
$\mu - \mu^3 = (2/5)\mathsf{P}_1(\mu) - (2/5)\mathsf{P}_3(\mu)$.

Using the above formula, angular integrations of
Eqs.~(\ref{eq:2-6-04a})--(\ref{eq:2-6-04c}) are evaluated as
\begin{widetext}

\begin{align}
&  \int \frac{d\varOmega_{\bm{p}}}{4\pi}
  \left|\bar{\bm{L}}_2(\bm{k},\bm{p})\right|^2
  = 
   \frac{9}{49}
   W^2(kR) W^2(pR)\,
    \int_0^\infty r\,dr
  \left[ \frac{8}{15} j_0(kr) j_0(pr)
    - \frac{16}{21} j_2(kr) j_2(pr)
    + \frac{8}{35} j_4(kr) j_4(pr)
  \right],
  \label{eq:2-6-07a}\\
&  \int \frac{d\varOmega_{\bm{p}}}{4\pi}\,
  \bm{k}\cdot\bar{\bm{L}}_2(\bm{k},\bm{p})
  = \frac{3}{7} W(kR) W(pR)
  \int r\,dr
  \Biggl\{
    \frac{2}{3}k^2
    \left[
      j_0(kr) j_0(pr) - j_2(kr) j_2(pr)
    \right]
    + \frac{2}{5} kp
    \left[
      j_3(kr) j_3(pr) - j_1(kr) j_1(pr)
    \right]
  \Biggr\},
  \label{eq:2-6-07b}\\
&  \int \frac{d\varOmega_{\bm{p}}}{4\pi}\,
  \bm{k}\cdot\bar{\bm{L}}_3(\bm{k},\bm{p},-\bm{p})
  =
  \frac{10}{21} W(kR) W^2(pR) \, k^2
  \int_0^\infty r\,dr
  \left[
    \frac{8}{15} j_0(kr) j_0(pr)
    - \frac{16}{21} j_2(kr) j_2(pr)
    + \frac{8}{35} j_4(kr) j_4(pr)
  \right],
  \label{eq:2-6-07c}
\end{align}
where the first smoothing scheme, Eq.~(\ref{eq:2-2-04a}) is adopted.
Substituting the above results into Eq.~(\ref{eq:2-6-03}), we have
\begin{multline}
  \frac{\sigma_\mathrm{nv}^2}{a^2H^2f^2} = 
  \sigma_\mathrm{dpk}^2 + 
  \frac{18}{49} \int_0^\infty r\,dr
  \left\{
    \frac{8}{15}\left[\bar{\xi}^{(0)}_0(r)\right]^2
    - \frac{16}{21}\left[\bar{\xi}^{(0)}_2(r)\right]^2
    + \frac{8}{35}\left[\bar{\xi}^{(0)}_4(r)\right]^2
  \right\}
  \\
  + \frac{12}{7} \int_0^\infty r\,dr
  \left\{
    \frac{2}{3}
    \left[
      \bar{A}^{(0)}_0(r) \bar{B}^{(0)}_0(r)
      - \bar{A}^{(0)}_2(r) \bar{B}^{(0)}_2(r)
    \right]
    + \frac{2}{5}
    \left[
      \bar{A}^{(-1)}_3(r) \bar{B}^{(1)}_3(r)
      - \bar{A}^{(-1)}_1(r) \bar{B}^{(1)}_1(r)
    \right]
  \right\}
  \\
  + \frac{30}{21} \int_0^\infty r\,dr
  \left\{
    \frac{8}{15} \bar{A}^{(0)}_0(r) \bar{\xi}^{(0)}_0(r)
    - \frac{16}{21} \bar{A}^{(0)}_2(r) \bar{\xi}^{(0)}_2(r)
    + \frac{8}{35} \bar{A}^{(0)}_4(r) \bar{\xi}^{(0)}_4(r)
  \right\},
  \label{eq:2-6-08}
\end{multline}
\end{widetext}
where
\begin{align}
  \bar{\xi}^{(n)}_l(r) &\equiv
  \int_0^\infty\frac{k^2dk}{2\pi^2}\, k^n j_l(kr) W^2(kR) P_\mathrm{L}(k),
  \label{eq:2-6-09a}\\
  \bar{A}^{(n)}_l(r) &\equiv
  \int_0^\infty\frac{k^2dk}{2\pi^2}\, k^n
  \left(1 - \frac{1}{2} {R_\mathrm{v}}^2 k^2\right) j_l(kr)
  W^2(kR) P_\mathrm{L}(k)
  \nonumber\\
  &= \bar{\xi}^{(n)}_l(r)
  - \frac{1}{2} {R_\mathrm{v}}^2 \bar{\xi}^{(n+2)}_l(r),
  \label{eq:2-6-09b}\\
  \bar{B}^{(n)}_l(r) &\equiv
  \int_0^\infty\frac{k^2dk}{2\pi^2}\, k^n
  c^{(1)}_X(k) j_l(kr) W(kR) P_\mathrm{L}(k)
  \nonumber\\
  &= b_{10} \bar{\xi}^{(n)}_l(r) + b_{01} \bar{\xi}^{(n+2)}_l(r).
  \label{eq:2-6-09c}
\end{align}


The number-weighted dispersion of displacement field,
$\sigma_\mathrm{nd}^2 \equiv \langle
(1+\delta^\mathrm{L}_X)|\bm{\varPsi}|^2\rangle$, is just obtained by
replacements $\bar{\bm{L}}_n \rightarrow \bar{\bm{L}}_n/naHf$ in the
expression of the number-weighted dispersion of the velocity field.
Corresponding to Eq.~(\ref{eq:2-6-04}), we have
\begin{multline}
    \sigma_\mathrm{nd}^2 =
    \sigma_\mathrm{dpk}^2 + 
    \int\frac{d^3k}{(2\pi)^3} \frac{d^3p}{(2\pi)^3}
    P_\mathrm{L}(k) P_\mathrm{L}(p)
    \\ \times
    \Biggl\{
      \frac{1}{2} \left| \bar{\bm{L}}_2(\bm{k},\bm{p}) \right|^2
      + \frac{W(kR)}{k^2}
      \left(1 - \frac{1}{2}{R_\mathrm{v}}^2k^2\right)
    \\ \times
      \left[
        2 c^{(1)}_X(p)\,\bm{k}\cdot\bar{\bm{L}}_2(\bm{k},\bm{p}) +
        \bm{k}\cdot\bar{\bm{L}}_3(\bm{k},\bm{p},-\bm{p})
      \right]
    \Biggr\}.
    \label{eq:2-6-11}
\end{multline}
Correspondingly, Eq.~(\ref{eq:2-6-08}) can easily be modified to give
the formula for $\sigma_{\mathrm{nd}}^2$ by replacing the coefficients
in front of three integrals as $18/49 \rightarrow 9/98$,
$12/7 \rightarrow 6/7$ and $30/21 \rightarrow 10/21$.

\section{\label{sec:General} General formulation of velocity bias with
  the flat constraint}

\subsection{\label{subsec:}
A class of bias models with the flat constraint
}

The appearance of the linear velocity bias from a second term in
Eq.~(\ref{eq:2-2-01}) is due to the $\chi_1$ term in the second-order
renormalized bias function of Eq.~(\ref{eq:2-2-07b}), which is an odd
function of the wave vector $\bm{p}$. The same is also the case for the
linear bias factor of density field in Eq.~(\ref{eq:2-3-3}). Thus it
is crucial that the $\chi_1$ term is present in the renormalized bias
function to have the effects of velocity bias.

The $\chi_1$ factor arises from the flat constraint of peaks,
$\bm{\eta} = \bm{0}$, where $\bm{\eta} = \bm{\nabla}\delta_R/\sigma_1$
\cite{Des13,LMD15,MD16}. Therefore, the flat constraint is an
essential ingredient for the appearance of velocity bias. This is
simply interpreted as follows: at a point with the flat constraint,
density gradients are zero, and the magnitude of velocity at the same
point is expected to be smaller than the average value. Thereby the
velocity is statistically biased in the presence of the flat
constraint.

The peak constraint contains the flat constraint, and the velocity
bias appears from the $\chi_1$ term in the lowest-order approximation
in the last section. Any other model with the flat constraint, such as
the excursion set peaks (ESP) \cite{AP90,PS12}, is naturally expected
to have the velocity bias of the same kind. In the following, we
consider a general situation in the presence of the flat constraint,
in order to investigate the origin and properties of the velocity
bias.

For the sake of generality, we assume a general constraint in the
Lagrangian number density of objects which contains the flat
constraint,
\begin{equation}
  n_X(\bm{x})
  = \left(\frac{2\pi}{3}\right)^{3/2}
  F(\nu,\bm{\zeta},\ldots)\,
  \delta_\mathrm{D}^3(\bm{\eta}).
  \label{eq:3-1-01}
\end{equation}
In the above notation, $\nu$ is a scalar, $\bm{\eta}$ is a
three-dimensional vector, and $\bm{\zeta}$ is a $3\times 3$ tensor,
which components are defined by
\begin{equation}
  \label{eq:3-1-02}
  \nu(\bm{x}) = \frac{\delta_R(\bm{x})}{\sigma_0},\quad
  \eta_i(\bm{x}) = \frac{\partial_i\delta_R(\bm{x})}{\sigma_1},\quad
  \zeta_{ij}(\bm{x})
  = \frac{\partial_i\partial_j\delta_R(\bm{x})}{\sigma_2}.
\end{equation}
The function $F$ is an arbitrary function of the above variables at a
location $\bm{x}$ where the number density of biased objects
$n_X(\bm{x})$ is defined. From the rotational symmetry, the variable
$\bm{\eta}$ is uncorrelated with variables $\nu$ and $\bm{\zeta}$:
$\langle \nu \eta_i \rangle = \langle \eta_i \zeta_{jk} \rangle = 0$.
We assume the function $F$ can depend on other variables which are
assumed to be uncorrelated with $\bm{\eta}$. For example, the variable
$\bm{\eta}$ is correlated with the third-order derivatives
$\partial_i\partial_j\partial_k\delta_R$, which is assumed to be
absent in the function $F$. The ESP model has an additional variable
$\partial\delta_R/\partial R$, and this variable is really
uncorrelated with $\bm{\eta}$ because of the rotational symmetry. Thus
the ESP model is an example with the general constraint of
Eq.~(\ref{eq:3-1-01}).

In the general formalism below, different smoothing kernels can be
adopted for every variable, $\nu$, $\bm{\eta}$, $\bm{\zeta}$,
$\ldots$. For example, both the top-hat kernel and the Gaussian kernel
are adopted in a certain version of the ESP model. The essential
assumption below is that $\bm{\eta}$ is the only variable which has
odd parity. Other variables are inevitably uncorrelated to $\bm{\eta}$
at a single point by parity symmetry.

\subsection{\label{subsec:ImplicationFlat}
Implications of the flat constraint
and the renormalized bias functions
}

The observable quantities, such as the power spectrum, depends on the
series of renormalized bias functions. The renormalized bias function
$c_X^{(n)}$ is defined by \cite{Mat12}
\begin{equation}
  \left\langle
      \frac{\delta^n \delta_X^\mathrm{L}(\bm{k})}
      {\delta\delta_{\rm L}(\bm{k}_1)
        \cdots\delta\delta_{\rm L}(\bm{k}_n)}
  \right\rangle = 
  (2\pi)^{3-3n}\delta_{\rm D}^3(\bm{k}-\bm{k}_{1\cdots n})
  c^{(n)}_X(\bm{k}_1,\ldots,\bm{k}_n),
  \label{eq:3-2-01}
\end{equation}
where $\delta_X^\mathrm{L}(\bm{k})$ is the Fourier transform of the
density contrast of the biased objects in Lagrangian space, and
$\bm{k}_{1\cdots n} \equiv \bm{k}_1+\cdots +\bm{k}_n$. The density
contrast in Lagrangian space is given by $n_X(\bm{x})/\bar{n}_X-1$,
where $\bar{n}_X = \langle n_X \rangle$ is the mean number density of
biased objects.

In a class of models in which the number density $n_X(\bm{x})$ is
given by a function of finite number of variables, $y_\alpha(\bm{x})$,
which are linearly related to the linear density field
$\delta_\mathrm{L}$, the renormalized bias function in
Eq.~(\ref{eq:3-2-01}) is given by \cite{Mat11}
\begin{multline}
  \label{eq:3-2-02}
  c_X^{(n)}(\bm{k}_1,\ldots,\bm{k}_n) =
  \frac{1}{\bar{n}_X}
  \sum_{\alpha_1,\ldots,\alpha_n}
  \left\langle
    \frac{\partial^n n_X}
    {\partial y_{\alpha_1} \cdots \partial y_{\alpha_n}}
  \right\rangle
  \\ \times
  U_{\alpha_1}(\bm{k}_1) \cdots U_{\alpha_n}(\bm{k}_n),
\end{multline}
where $U_\alpha(\bm{k})$ are the Fourier coefficients of the variables
$y_\alpha(\bm{x})$,
\begin{equation}
  \label{eq:3-2-03}
  y_\alpha(\bm{x}) = 
  \int \frac{d^3k}{(2\pi)^3} e^{i\bm{k}\cdot\bm{x}}
  U_\alpha(\bm{k}) \delta_\mathrm{L}(\bm{k}).
\end{equation}
Defining an operator
\begin{equation}
  \hat{D}(\bm{k}) = \sum_\alpha U_\alpha(\bm{k})
  \frac{\partial}{\partial y_\alpha}\,,
  \label{eq:3-2-04}
\end{equation}
Eq.~(\ref{eq:3-2-02}) can be represented by
\begin{equation}
  c_X^{(n)}(\bm{k}_1,\ldots,\bm{k}_n) =
  \frac{1}{\bar{n}_X}
  \left\langle
    \hat{D}(\bm{k}_1) \cdots \hat{D}(\bm{k}_n) n_X
  \right\rangle.
  \label{eq:3-2-05}
\end{equation}

In Eqs.~(\ref{eq:3-2-02}) and (\ref{eq:3-2-05}), the average
$\langle\cdots\rangle$ is taken over the random variables
$\bm{y}=(y_\alpha)$. We assume Gaussian initial conditions throughout
this paper, and the distribution function is given by
\cite{Dor70,BBKS,PGP09,GPP12,Des13,LMD15}
\begin{equation}
  P(\bm{y}) = N_0
  \exp\left[
    - \frac{\nu^2 + {J_1}^2 - 2\gamma \nu J_1}{2(1-\gamma^2)}
    - \frac{3}{2} \eta^2 - \frac{5}{2} J_2
  \right].
  \label{eq:3-2-06}
\end{equation}
In the above equation, we introduce notations
\begin{equation}
  \label{eq:3-2-07}
  \eta^2 \equiv \bm{\eta}\cdot\bm{\eta}, \quad
  J_1 \equiv - \zeta_{ii}, \quad 
  J_2 \equiv \frac{3}{2} \tilde{\zeta}_{ij} \tilde{\zeta}_{ji}
\end{equation}
where repeated indices are summed over, and
\begin{equation}
  \label{eq:3-2-08}
  \tilde{\zeta}_{ij} \equiv \zeta_{ij} + \frac{1}{3}\delta_{ij} J_1
\end{equation}
is the traceless part of $\zeta_{ij}$. The factor $N_0$ is a
normalization constant to ensure that the total probability is equal
to one, but its actual value is not used for our applications in this
paper.

Although the RHS of Eq.~(\ref{eq:3-2-06}) is represented by
rotationally invariant variables $\nu$, $J_1$ and $J_2$, the
probability distribution function $P(\bm{y})$ is still for the linear
variables $\bm{y}$, and not for the invariant variables.

In our class of models, Eq.~(\ref{eq:3-1-01}), the variables
$y_\alpha$ are given by
\begin{equation}
   \left(y_\alpha\right) =
   \left(
     \nu, \eta_i, \zeta_{ij},
     \ldots
   \right),
  \label{eq:3-2-09}
\end{equation}
and
\begin{equation}
  \left[U_\alpha(\bm{k})\right] =
  \left(
    \frac{W(kR)}{\sigma_0},
    \frac{ik_iW(kR)}{\sigma_1},
    -\frac{k_ik_jW(kR)}{\sigma_2},
    \ldots
  \right),
  \label{eq:3-2-10}
\end{equation}
where $i\leq j$. When different smoothing functions are applied to
each variable, the window function $W(kR)$ in the above equation is
replaced by corresponding functions. Since we assume the variable
$\bm{\eta}$ is uncorrelated to other variables, the only components
which have odd parity, $U_\alpha(-\bm{k}) = - U_\alpha(\bm{k})$, are
the components $ik_iW(kR)/\sigma_1$ in Eq.~(\ref{eq:3-2-10}). All the
other components are assumed to have even parity,
$U_\alpha(-\bm{k}) = + U_\alpha(\bm{k})$.

It is an essential assumption in the bias models analyzed in this
paper that the variable $\bm{\eta}$ is the only quantity which has odd
parity in the set of variables $(y_\alpha)$. This is the case for the
peaks model and is the reason why only the term with $\chi_1$ in
$c_X^{(2)}$ survives and causes the velocity bias as observed in
Sec.~\ref{sec:LinearVelocityBias}. It is essential that the term with
$\chi_1$ in $c_X^{(2)}(\bm{k},\bm{p})$ of Eq.~(\ref{eq:2-2-07b}) is
the only term which is an odd function of $\bm{p}$. All the other terms
are even functions of $\bm{p}$, so that they vanish in the angular
integral of $c_X^{(2)}(\bm{k},\bm{p})\bm{L}_1(-\bm{p})$. In order to
give a general formalism in this section, we pursue similar mechanisms
in higher-order terms.

For that purpose, we decompose the operator $\hat{D}(\bm{k})$ of
Eq.~(\ref{eq:3-2-04}) as
\begin{equation}
  \hat{D}(\bm{k}) =
  \hat{D}_0(\bm{k}) + \hat{D}_1(\bm{k})\,,
  \label{eq:3-2-11}
\end{equation}
where
\begin{equation}
  \hat{D}_0(\bm{k})
  = \frac{W(kR)}{\sigma_0} \frac{\partial}{\partial\nu}
  - \frac{W(kR)}{\sigma_2} \sum_{i\leq j} k_i k_j 
  \frac{\partial}{\partial\zeta_{ij}} + \cdots
  \label{eq:3-2-12}
\end{equation}
corresponds to an operator of even parity, $\hat{D}_0(-\bm{k}) =
+\hat{D}_0(\bm{k})$, and
\begin{equation}
  \hat{D}_1(\bm{k}) = 
  \frac{i W(kR)}{\sigma_1} \sum_i k_i
  \frac{\partial}{\partial\eta_i}
  \label{eq:3-2-13}
\end{equation}
corresponds to an operator of odd parity, $\hat{D}_1(-\bm{k}) =
-\hat{D}_1(\bm{k})$.

\begin{figure*}
\begin{center}
\includegraphics[width=40pc]{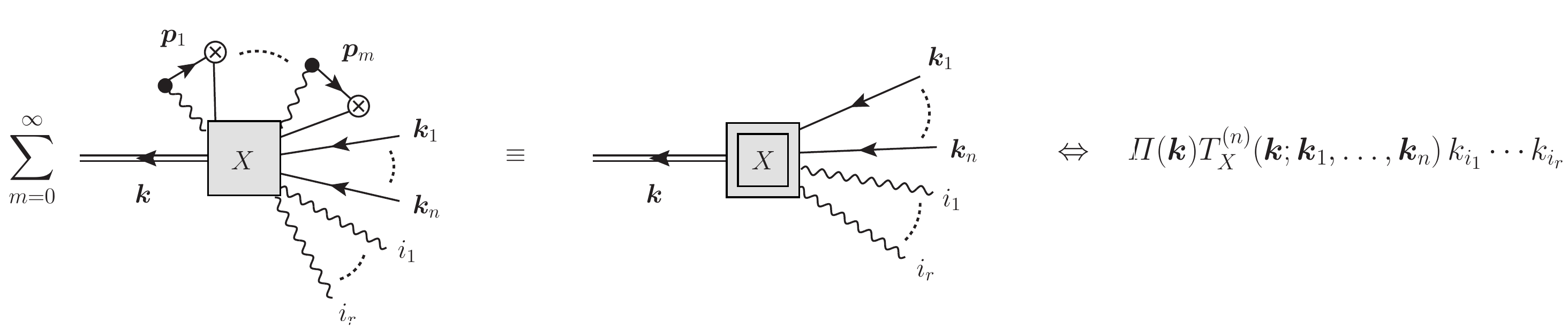}
\caption{\label{fig:SumSquare} A further resummation of the iPT vertex
  in the presence of flat constraint. }
\end{center}
\end{figure*}
We consider the following sum of integrals:
\begin{multline}
  T_X^{(n)}(\bm{k};\bm{k}_1,\ldots,\bm{k}_n)
  \\
  \equiv
  \sum_{m=0}^\infty \frac{1}{m!}
  \int \frac{d^3p_1}{(2\pi)^3} \cdots \frac{d^3p_m}{(2\pi)^3}\,
  c_X^{(n+m)}(\bm{k}_1,\ldots,\bm{k}_n,\bm{p}_1,\ldots,\bm{p}_m)
\\
  \times
  \left[\bm{k}\cdot\bar{\bm{L}}_1(-\bm{p}_1)\right] \cdots
  \left[\bm{k}\cdot\bar{\bm{L}}_1(-\bm{p}_m)\right]
  P_\mathrm{L}(p_1)\cdots  P_\mathrm{L}(p_m).
  \label{eq:3-2-14}
\end{multline}
This form of integral appears in the applications of iPT, and a
corresponding section of the iPT diagram is given in the left diagram
of Fig.~\ref{fig:SumSquare}. For the diagrammatic rules of iPT, see
Refs.~\cite{Mat11,Mat14}. The corresponding iPT diagram is denoted by
a double square as shown in Fig.~\ref{fig:SumSquare}. This iPT diagram
has another resummation factor $\varPi(\bm{k}) k_{i_1}\cdots k_{i_m}$
in addition to the function $T_X^{(n)}$, where $\varPi(\bm{k})$ is the
resummation factor of the displacement field%
\footnote{The resummation of the iPT explicitly violates the
  ``extended Galilean invariance'' \cite{PP16} in the small-scale
  limit. Since the iPT is the theory for the weakly nonlinear regime
  and should not be applied on the scales below the shell-crossing
  scale, the violation is not considered to be a practical issue of
  iPT.} %
\cite{Mat11}. The exact diagrammatic rule for this vertex of the
double square is also shown in Fig.~\ref{fig:SumSquare}.

Because $\bm{L}_1(\bm{p}) = \bm{p}/p^2$ is an odd function of
$\bm{p}$, components of even parity in $c^{(n+m)}_X$ with respect to
$\bm{p}_1,\ldots,\bm{p}_m$ vanish. Substituting the expression of
Eq.~(\ref{eq:3-2-05}) into Eq.~(\ref{eq:3-2-14}),
$\hat{D}(\bm{p}_i)$'s are replaced by $\hat{D}_1(\bm{p}_i)$'s. Using
an identity,
\begin{multline}
  \int \frac{d^3p}{(2\pi)^3} \bar{\bm{L}}_1(-\bm{p})
  P_\mathrm{L}(p) \hat{D}(\bm{p})
  \\
  =
  - \int \frac{d^3p}{(2\pi)^3} \frac{\bm{p}}{p^2}
  W(pR) P_\mathrm{L}(p)  \hat{D}_1(\bm{p})
  = - \frac{i}{3} \frac{{\sigma_0}^2}{\sigma_1}
  \frac{\partial}{\partial\bm{\eta}}\,,
  \label{eq:3-2-15}
\end{multline}
we have
\begin{align}
  &  T_X^{(n)}(\bm{k};\bm{k}_1,\ldots,\bm{k}_n)
\nonumber\\    
  & \quad =  \frac{1}{\bar{n}_X} \sum_{m=0}^\infty
  \frac{1}{m!}
  \left(- \frac{i}{3} \frac{{\sigma_0}^2}{\sigma_1} \right)^m
  \left\langle
    \hat{D}(\bm{k}_1) \cdots \hat{D}(\bm{k}_n)
    \left(\bm{k}\cdot\frac{\partial}{\partial\bm{\eta}}\right)^m
    n_X
  \right\rangle
\nonumber\\
  &\quad =  \frac{1}{\bar{n}_X}
  \left\langle
    \hat{D}(\bm{k}_1) \cdots \hat{D}(\bm{k}_n)
    \exp\left(- \frac{i}{3} \frac{{\sigma_0}^2}{\sigma_1}
    \bm{k}\cdot\frac{\partial}{\partial\bm{\eta}}
  \right)
  n_X
  \right\rangle.
  \label{eq:3-2-16}
\end{align}
This expression reduces to a more useful form. In
Appendix~\ref{app:IntegralSummation}, it is shown that the function
$T_X^{(n)}(\bm{k};\ldots)$ generally has a factor
$\exp(k^2{\sigma_0}^4/6{\sigma_1}^2)$, and explicit forms for
$n=0,1,2$ are derived. We define the normalized function
$\hat{T}_X^{(n)}$ by
\begin{equation}
  T_X^{(n)}(\bm{k};\bm{k}_1,\ldots,\bm{k}_n) =
  \exp\left(\frac{k^2}{6}\frac{{\sigma_0}^4}{{\sigma_1}^2}\right)\,
  \hat{T}_X^{(n)}(\bm{k};\bm{k}_1,\ldots,\bm{k}_n),
  \label{eq:3-2-17}
\end{equation}
and the results for $n=0,1,2$ are given by Eqs.~(\ref{eq:a-7}),
(\ref{eq:a-10}), and (\ref{eq:a-13}). Substituting
$W(k_1R)\bm{k}_1 = {k_1}^2 \bar{\bm{L}}_1(\bm{k}_1)$,
$W(k_2R)\bm{k}_2 = {k_2}^2 \bar{\bm{L}}_1(\bm{k}_2)$ in these
equations, they can be represented by
\begin{align}
  \hat{T}_X^{(0)}(\bm{k})
  &= 1,
  \label{eq:3-2-18a}\\
  \hat{T}_X^{(1)}(\bm{k};\bm{k}_1)
  &=
  c_X^{(1)}(k_1)
  - {R_\mathrm{v}}^2 {k_1}^2 \bm{k}\cdot\bar{\bm{L}}_1(\bm{k}_1),
  \label{eq:3-2-18b}\\
  \hat{T}_X^{(2)}(\bm{k};\bm{k}_1,\bm{k}_2)
  &= c_X^{(2)}(\bm{k}_1,\bm{k}_2) - c_X^{(1)}(k_1) c_X^{(1)}(k_2)
    \nonumber \\
  &\hspace{5pc} +
  \hat{T}^{(1)}_X(\bm{k};\bm{k}_1)
  \hat{T}^{(1)}_X(\bm{k};\bm{k}_2).
  \label{eq:3-2-18c}
\end{align}

The above results are derived in real space. In redshift space, the
LPT kernels are replaced by \cite{Mat08}
\begin{equation}
  \bm{L}_n \rightarrow \bm{L}^\mathrm{s}_n =
  \bm{L}_n + nf\left(\hat{\bm{z}}\cdot\bm{L}_n\right)\hat{\bm{z}},
  \label{eq:3-2-19}
\end{equation}
where $\hat{\bm{z}}$ is the unit vector along the line of sight, and
$f=d\ln D/d\ln a$ is the linear growth rate. In particular, we have
\begin{equation}
  \bm{k}\cdot\bar{\bm{L}}_1(\bm{p})
  \rightarrow
  \bm{k}\cdot\bar{\bm{L}}^\mathrm{s}_1(\bm{p})
  = 
  \left[
    \bm{k} + fk_z\hat{\bm{z}}
  \right]
  \cdot\bar{\bm{L}}_1(\bm{p}),
  \label{eq:3-2-20}
\end{equation}
where $k_z = \bm{k}\cdot\hat{\bm{z}}$ is the line-of-sight component
of the vector $\bm{k}$. Since the $\bm{k}$-dependence and LPT kernels
in Eq.~(\ref{eq:3-2-14}) are included only in the form of
Eq.~(\ref{eq:3-2-20}), the function $T_X^{(n)}$ in redshift space is
given by Eq.~(\ref{eq:3-2-14}), with the replacement
$\bm{k} \rightarrow \bm{k} + fk_z\hat{\bm{z}}$, i.e.,
$T_X^{(n)}( \bm{k} + fk_z\hat{\bm{z}};\bm{k}_1,\ldots,\bm{k}_n)$. On
the one hand, this result is equivalent to just replacing the
first-order kernel $\bm{L}_1$ by $\bm{L}^\mathrm{s}_1$ in
Eqs.~(\ref{eq:3-2-18a})--(\ref{eq:3-2-18c}). On the other hand, $k^2$
is replaced by $k^2 + f(f+2){k_z}^2$ in the exponential factor of
Eq.~(\ref{eq:3-2-17}). Therefore, the function $T_X^{(n)}$ in redshift
space is given by
\begin{multline}
  T_X^{\mathrm{s}(n)}(\bm{k};\bm{k}_1,\ldots,\bm{k}_n) =
  \exp\left\{
    \frac{k^2}{6}\left[1+f(f+2){\mu_{\bm{k}}}^2\right]
    \frac{{\sigma_0}^4}{{\sigma_1}^2}
  \right\}
  \\ \times
  \hat{T}_X^{\mathrm{s}(n)}(\bm{k};\bm{k}_1,\ldots,\bm{k}_n),
  \label{eq:3-2-21}
\end{multline}
where $\mu_{\bm{k}} \equiv k_z/k$, and the function
$\hat{T}_X^{\mathrm{s}(n)}(\bm{k};\bm{k}_1,\ldots,\bm{k}_n)$ is given
by Eqs.~(\ref{eq:3-2-18a})--(\ref{eq:3-2-18c}) with a replacement
$\bar{\bm{L}}_1 \rightarrow \bar{\bm{L}}^\mathrm{s}_1$.

\subsection{\label{subsec:DensityProp}
Propagators of density field with the flat constraint
}

In this section, we consider propagators
$\varGamma_X^{(n)}(\bm{k}_1,\ldots,\bm{k}_n)$ of the density field
with the flat constraint in the formalism of iPT. Using the
propagators, the density power spectrum $P_X(\bm{k})$ of biased object
$X$ is given by
\begin{multline}
  P_X(\bm{k}) =
  \left[ \varGamma_X^{(1)}(\bm{k}) \right]^2 P_\mathrm{L}(k)
\\
  + \frac{1}{2}
  \int_{\bm{k}_1+\bm{k}_2 = \bm{k}}
  \left[ \varGamma_X^{(2)}(\bm{k}_1,\bm{k}_2) \right]^2
  P_\mathrm{L}(k_1) P_\mathrm{L}(k_2) + \cdots,
  \label{eq:3-3-11}
\end{multline}
in the case of the Gaussian initial condition. In more general cases
with primordial non-Gaussianity, other terms with the linear
bispectrum, trispectrum, etc., are added to the above expansion
\cite{Mat11,Mat12,YM13}. The effects of primordial non-Gaussianity
also modify the the calculations of the renormalized bias functions in
Sec.~\ref{subsec:ImplicationFlat}. In this paper, we only consider the
case of Gaussian initial conditions throughout.

\begin{figure*}
\begin{center}
\includegraphics[width=40pc]{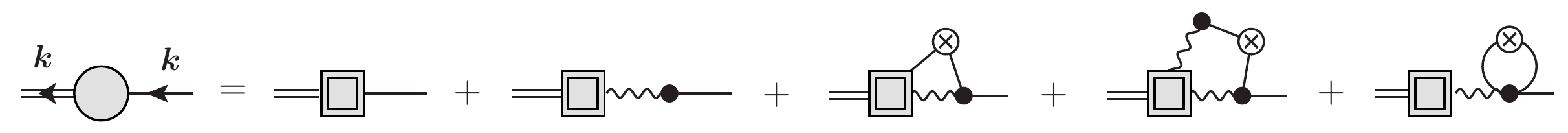}
\caption{\label{fig:PropOneS}
Two-point propagator with the resummed vertex.
}
\end{center}
\end{figure*}

In the course of evaluating the propagators, special types of loop
corrections, depicted by Fig.~\ref{fig:SumSquare} with the flat
constraint, are taken into account. First we consider the two-point
propagator. Using a diagrammatic notation of Fig.~\ref{fig:SumSquare}
for the iPT vertex, the two-point propagator,
$\varGamma^{(1)}_X(\bm{k})$, is given by Fig.~\ref{fig:PropOneS} up to
the one-loop approximation.
It is straightforward to apply the diagrammatic rules of iPT
\cite{Mat11} and a new rule of Fig.~\ref{fig:SumSquare} with the flat
constraint for a resummed vertex. The two-point propagator is
represented by $\varGamma_X^{(1)}(\bm{k}) =
\varPi(\bm{k})\hat{\varGamma}_X^{(1)}(\bm{k})$, where $\varPi(\bm{k})$
is the resummation factor of the displacement field, and
$\hat{\varGamma}_X^{(1)}(\bm{k})$ is the normalized two-point
propagator. The resummation factor in real space, $\varPi(k)$, is
given by \cite{Mat08,Mat11}
\begin{equation}
  \varPi(k) = \exp\left(-\frac{k^2}{6}{\sigma_{-1}}^2\right).
  \label{eq:3-3-1}
\end{equation}
up to the one-loop order.

Applying diagrammatic rules to Fig.~\ref{fig:PropOneS}, the normalized
two-point propagator up to the one-loop approximation is given by
\begin{multline}
  \hat{\varGamma}^{(1)}_X(\bm{k}) = 
  T_X^{(1)}(\bm{k};\bm{k})
  + T_X^{(0)}(\bm{k})\,\bm{k}\cdot\bar{\bm{L}}_1(\bm{k})
  \\
  + \int\frac{d^3p}{(2\pi)^3} P_\mathrm{L}(p)
  \Biggl\{
    \left[
      T_X^{(1)}(\bm{k};\bm{p})
      + T_X^{(0)}(\bm{k})\,\bm{k}\cdot\bar{\bm{L}}_1(\bm{p})
    \right]\,
    \bm{k}\cdot\bar{\bm{L}}_2(\bm{k},-\bm{p})
    \\
    + \frac{1}{2}
    T_X^{(0)}(\bm{k})\,\bm{k}\cdot\bar{\bm{L}}_3(\bm{k},\bm{p},-\bm{p})
  \Biggr\}.
  \label{eq:3-3-2}
\end{multline}
Because the function $T_X^{(n)}$ also has a
common factor as in Eq.~(\ref{eq:3-2-17}), the propagators in general
have a common factor
\begin{equation}
  G_\mathrm{d}(k) \equiv
  \varPi(k)\,
  \exp\left(\frac{k^2}{6}\frac{{\sigma_0}^4}{{\sigma_1}^2}\right)
  =
  \exp\left(
    -\frac{k^2}{6} \sigma^2_\mathrm{dpk}
  \right),
  \label{eq:3-3-3}
\end{equation}
where $\sigma_\mathrm{dpk}^2$ is the peak displacement dispersion
defined by Eq.~(\ref{eq:2-6-04}). Substituting
Eqs.~(\ref{eq:3-2-18a})--(\ref{eq:3-2-18c}) into Eq.~(\ref{eq:3-3-2}),
we obtain
\begin{align}
&  \frac{\varGamma_X^{(1)}(\bm{k})}{G_\mathrm{d}(k)} = 
  c_X^{(1)}(k) +
  \left(1-{R_\mathrm{v}}^2 k^2\right)\,
  \bm{k}\cdot\bar{\bm{L}}_1(\bm{k})
\nonumber\\
& \quad
  + \int\frac{d^3p}{(2\pi)^3} P_\mathrm{L}(p)
\nonumber\\
& \qquad \times
  \Biggl\{
    \left[
      c_X^{(1)}(p) +
      \left(1 - {R_\mathrm{v}}^2 p^2\right)
      \bm{k}\cdot\bar{\bm{L}}_1(\bm{p})
    \right]\,
    \bm{k}\cdot\bar{\bm{L}}_2(\bm{k},-\bm{p})
  \nonumber\\       
  & \hspace{10pc}
    + \frac{1}{2}
    \bm{k}\cdot\bar{\bm{L}}_3(\bm{k},\bm{p},-\bm{p})
  \Biggr\}.
  \label{eq:3-3-5}
\end{align}
Comparing the above equation with Eq.~(\ref{eq:2-3-1}), the
resummation of the flat constraint in the bias replaces the
first-order kernel $\bar{\bm{L}}_1(\bm{k})$ by
$(1-{R_\mathrm{v}}^2k^2)\bar{\bm{L}}_1(\bm{k})$, removes the
second-order bias function $c^{(2)}_X$ in the integrand of the
one-loop term, and converts $\sigma_{-1}^2$ into
$\sigma_\mathrm{dpk}^2$ in the exponential prefactor.

Neglecting mode-coupling terms in
Eq.~(\ref{eq:3-3-5}), the effective bias factor in Eulerian space is
given by
\begin{align}
  \left.\varGamma_X^{(1)}(k)\right|_\mathrm{tree}
  &= G_\mathrm{d}(k)
  \left[
    c_X^{(1)}(k) + \left(1-{R_\mathrm{v}}^2 k^2\right) W(kR)
  \right]
\nonumber\\
   &= G_\mathrm{d}(k)
  \left[
    b_\mathrm{v}(k) + c_X^{(1)}(k)
  \right]
  \equiv b_X^\mathrm{eff}(k).
  \label{eq:3-3-6}
\end{align}
The two-point propagator $\varGamma_X^{(1)}(k)$ is effectively a bias
factor $b_X^\mathrm{eff}(k)$ in real space. This result agrees with a
previous result, Eq.~(10) of Ref.~\cite{Bal15} in the peaks model.

When the factor $\exp(k^2{\sigma_0}^4/6{\sigma_1}^2)$ is dropped, this
result is consistent with that of the previous section,
Eq.~(\ref{eq:2-3-1}). The reason for not showing this factor in the
previous section is that this contribution comes from two- and
higher-loop corrections in terms of the original diagram of iPT. As
one can explicitly show, the diagram of Fig.~\ref{fig:SumSquare} with
even numbers $m=2l$ gives a factor of
$(k^2{\sigma_0}^4/6{\sigma_1}^2)^l/l!$. The diagram with odd numbers
$m=2l+1$ only contributes when $n \geq 1$, and gives a factor of
${R_\mathrm{v}}^2 (k^2{\sigma_0}^4/6{\sigma_1}^2)^l/l!$. After the
summation over $l$, the factor $\exp(k^2{\sigma_0}^4/6{\sigma_1}^2)$
or ${R_\mathrm{v}}^2\exp(k^2{\sigma_0}^4/6{\sigma_1}^2)$ emerges. If
we take only the cases $m=0$ and $1$ into account, the exponential
factor does not appear. Since $m$ is the number of loops in the
diagram of Fig.~\ref{fig:SumSquare}, the exponential factor comes from
two-, or higher-loop corrections. In this way, loop corrections with
the flat constraint contains contributions without mode couplings,
which look like a linear term.

The above results are explicitly derived for real space. As noted in
the previous subsection, extending the results to redshift space is
straightforward. In redshift space, the resummation factor
$\varPi(\bm{k})$ is given by \cite{Mat08,Mat11}
\begin{equation}
  \varPi^\mathrm{s}(\bm{k}) =
  \exp\left\{
    - \frac{k^2}{6}\left[1+f(f+2){\mu_{\bm{k}}}^2\right]{\sigma_{-1}}^2
  \right\},
  \label{eq:3-3-7}
\end{equation}
while the exponential factor in $T_X^{(n)}$ is given by
Eq.~(\ref{eq:3-2-21}). Accordingly, the factor $G_\mathrm{d}(\bm{k})$
of Eq.~(\ref{eq:3-3-3}) in real space is replaced by
\begin{equation}
  G_\mathrm{d}^\mathrm{s}(\bm{k}) \equiv
  \exp\left\{
    - \frac{k^2}{6}\left[1+f(f+2){\mu_{\bm{k}}}^2\right]
    \sigma^2_\mathrm{dpk}
  \right\},
  \label{eq:3-3-8}
\end{equation}
in redshift space.
The two-point propagator in redshift space
$\varGamma^{\mathrm{s}(n)}_X$ is given by the form of
Eq.~(\ref{eq:3-3-5}) with replacements $\bar{\bm{L}}_n \rightarrow
\bar{\bm{L}}^\mathrm{s}_n$ and $G_\mathrm{d}(k) \rightarrow
G_\mathrm{d}^\mathrm{s}(\bm{k})$.

In redshift space, the corresponding factor of Eq.~(\ref{eq:3-3-6}) is
given by
\begin{equation}
  \left.\varGamma_X^{\mathrm{s}(1)}(\bm{k})\right|_\mathrm{tree}
  = G^\mathrm{s}_\mathrm{d}(\bm{k})
  \left[
    b_\mathrm{v}(k)
    \left( 1 +  f {\mu_{\bm{k}}}^2 \right)
    + c_X^{(1)}(k)
  \right].
  \label{eq:3-3-12}
\end{equation}
In the large-scale limit, $k\rightarrow 0$, this expression reduces to
the Kaiser's factor, $b+f{\mu_{\bm{k}}}^2=b(1+f{\mu_{\bm{k}}}^2/b)$,
for the linear redshift-space distortion \cite{Kai87}. It is a common
practice to define the redshift-space distortion parameter $\beta =
f/b$ in the Kaiser's limit. In Eq.~(\ref{eq:3-3-12}), this parameter
is effectively scale dependent. The corresponding parameter is given
by
\begin{equation}
  \beta_\mathrm{eff}(k) \equiv 
  G_\mathrm{d}(k)
  b_\mathrm{v}(k)
  \frac{f}{b_X^\mathrm{eff}(k)},
  \label{eq:3-3-13}
\end{equation}
and Eq.~(\ref{eq:3-3-12}) is represented by
\begin{multline}
  \left.\varGamma_X^{\mathrm{s}(1)}(\bm{k})\right|_\mathrm{tree}
  =
  \exp\left[
    - \frac{k^2}{6}f(f+2)\sigma^2_\mathrm{dpk}{\mu_{\bm{k}}}^2
  \right]
  \\
  \times
  b^\mathrm{eff}_X(k)
  \left[
    1 + \beta_\mathrm{eff}(k) {\mu_{\bm{k}}}^2
  \right].
  \label{eq:3-3-14}
\end{multline}
The exponential prefactor corresponds to the damping factor along the
line of sight, which represents the fingers-of-God effect of biased
objects within the Zel'dovich approximation.

\begin{figure*}
\begin{center}
\includegraphics[width=40pc]{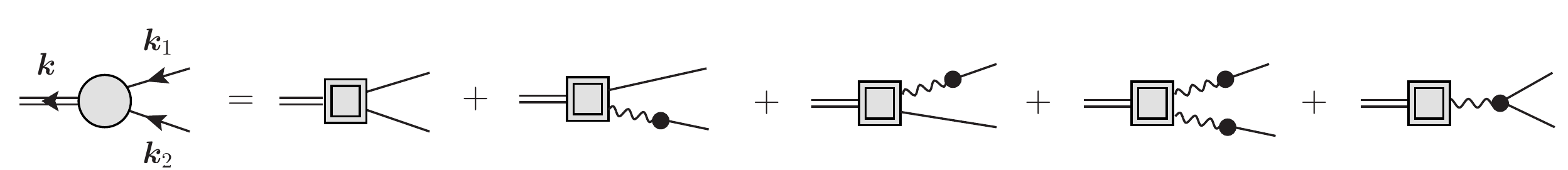}
\caption{\label{fig:PropTwoS}
Three-point propagator with the resummed vertex.
}
\end{center}
\end{figure*}

Next we consider the three-point propagator,
$\varGamma_X^{(2)}(\bm{k}_1,\bm{k}_2) =
\varPi(\bm{k}_{12})\hat{\varGamma}_X^{(2)}(\bm{k}_1,\bm{k}_2)$, where
$\bm{k}_{12} = \bm{k}_1 + \bm{k}_2$. The diagram for this propagator
is given in Fig.~\ref{fig:PropTwoS}.
In calculating the power spectrum, the three-point propagator is
always accompanied by loop diagrams. Thus, we do not need to consider
loop diagrams for the three-point propagator for the purpose of
evaluating the power spectrum in the one-loop approximation. Applying
diagrammatic rules to Fig.~\ref{fig:PropTwoS}, the normalized
three-point propagator is given by
\begin{multline}
  \hat{\varGamma}^{(2)}_X(\bm{k}_1,\bm{k}_2) = 
  T_X^{(2)}(\bm{k};\bm{k}_1,\bm{k}_2)
  + T_X^{(1)}(\bm{k};\bm{k}_1)\,\bm{k}\cdot\bar{\bm{L}}_1(\bm{k}_2)
\\
  + T_X^{(1)}(\bm{k};\bm{k}_2)\,\bm{k}\cdot\bar{\bm{L}}_1(\bm{k}_1)
  + T_X^{(0)}(\bm{k})\left[\bm{k}\cdot\bar{\bm{L}}_1(\bm{k}_1)\right]
    \left[\bm{k}\cdot\bar{\bm{L}}_1(\bm{k}_2)\right]
\\
  + T_X^{(0)}(\bm{k})\,
    \bm{k}\cdot\bar{\bm{L}}_2(\bm{k}_1,\bm{k}_2),
  \label{eq:3-3-9}
\end{multline}
where $\bm{k}=\bm{k}_1+\bm{k}_2$. Substituting
Eqs.~(\ref{eq:3-2-18a})--(\ref{eq:3-2-18c}), we have
\begin{multline}
  \frac{\varGamma_X^{(2)}(\bm{k}_1,\bm{k}_2)}{G_\mathrm{d}(k)} = 
  c_X^{(2)}(\bm{k}_1,\bm{k}_2)
\\
  +
  \left[
    c_X^{(1)}(k_1)
    \left(1-{R_\mathrm{v}}^2 {k_2}^2\right)\,
    \bm{k}\cdot\bar{\bm{L}}_1(\bm{k}_2)
    + (\bm{k}_1 \leftrightarrow \bm{k}_2)
  \right]
\\
  + \left(1-{R_\mathrm{v}}^2 {k_1}^2\right)
  \left(1-{R_\mathrm{v}}^2 {k_2}^2\right)
    \left[\bm{k}\cdot\bar{\bm{L}}_1(\bm{k}_1)\right]
    \left[\bm{k}\cdot\bar{\bm{L}}_1(\bm{k}_2)\right]
    \\
    + \bm{k}\cdot\bar{\bm{L}}_2(\bm{k}_1,\bm{k}_2).
  \label{eq:3-3-10}
\end{multline}
Comparing the above equation with Eq.~(16) of Ref.~\cite{Mat14}, the
resummation of the flat constraint in the bias replaces the
first-order kernel $\bar{\bm{L}}_1(\bm{k})$ by
$(1-{R_\mathrm{v}}^2k^2)\bar{\bm{L}}_1(\bm{k})$ and converts
$\sigma_{-1}^2$ into $\sigma_\mathrm{dpk}^2$ in the exponential
prefactor. The redshift-space counterpart is again obtained by
replacements, $\bar{\bm{L}}_n \rightarrow \bar{\bm{L}}^\mathrm{s}_n$
and $G_\mathrm{d}(k) \rightarrow G_\mathrm{d}^\mathrm{s}(\bm{k})$, in
the above expression.

Substituting Eqs.~(\ref{eq:3-3-5}) and (\ref{eq:3-3-10}) into
Eq.~(\ref{eq:3-3-11}), the expression for the power spectrum
$P_X(\bm{k})$ is obtained. In the case of Zel'dovich approximation,
$\bar{\bm{L}}_2 = \bar{\bm{L}}_3 = \cdots = 0$, the resulting
expression is consistent with a previous result of Ref.~\cite{BD17} in
the peaks model.

\subsection{\label{subsec:VelocityProp}
Propagators of momentum field with the flat constraint
}

\begin{figure}
\begin{center}
\includegraphics[width=19pc]{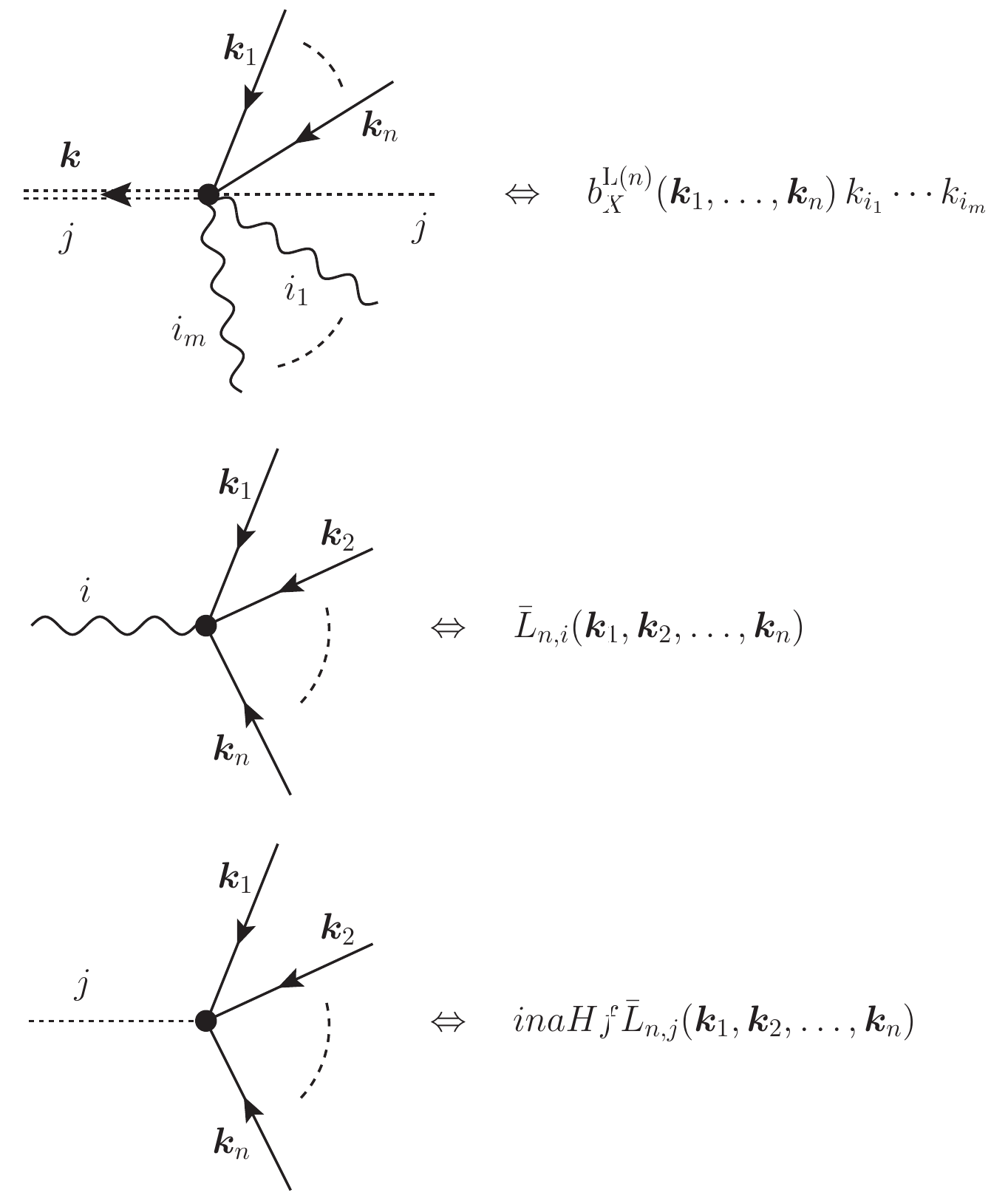}
\caption{\label{fig:MomRules} Diagrammatic rules for the momentum
  field $\bm{j}_X$. The top graph corresponds to an external vertex,
  and the middle and bottom graphs correspond to internal vertices.
  The symbols $k_i$ and $L_{n,i}$ are $i$ components of $\bm{k}$ and
  $\bm{L}_n$, respectively. The wavy lines and the single dashed lines
  are internal, while the double dashed lines are external. Every line
  carries wave vectors, and the sum of the wave vectors at each vertex
  should vanish. Only one single dotted line and only one double
  dotted line should be attached to the external vertex.}
\end{center}
\end{figure}

Extending the method of the previous subsection, we consider the
propagators of the velocity field in this section. For this purpose,
we need to have the diagrammatic rules of iPT for the velocity
momentum, $\bm{j}_X$. One can straightforwardly follow the derivation
of Ref.~\cite{Mat11} for the diagrammatic rules of iPT. The Fourier
transform of the momentum field $\bm{j}_X(\bm{x})$,
Eq.~(\ref{eq:2-1-03}), is given by
\begin{multline}
  \tilde{\bm{j}}_X(\bm{k}) =
  \sum_{m=0}^\infty \frac{(-i)^m}{m!}
  \sum_{n=0}^\infty \frac{1}{n!}
  \int_{\bm{k}'_{1\cdots n} + \bm{k}''_{1\cdots m} +
    \bm{k}'''=\bm{k}} 
  b_X^{\mathrm{L}(n)}(\bm{k}'_1,\ldots,\bm{k}'_n)
  \\ \times
  \delta_\mathrm{L}(\bm{k}'_1)\cdots\delta_\mathrm{L}(\bm{k}'_n)
  \left[
    \bm{k}\cdot\tilde{\bm{\varPsi}}(\bm{k}''_1)
  \right] \cdots
  \left[
    \bm{k}\cdot\tilde{\bm{\varPsi}}(\bm{k}''_m)
  \right]
  a \dot{\tilde{\bm{\varPsi}}}(\bm{k}'''),
  \label{eq:3-4-01}
\end{multline}
where the expansion of the number density field,
Eq.~(\ref{eq:2-2-05a}) is applied. The expansion of the displacement
field $\tilde{\bm{\varPsi}}(\bm{k})$ is given by
Eq.~(\ref{eq:2-2-04}). From these expansions, the diagrammatic rules
for the momentum field are given by Fig.~\ref{fig:MomRules}.

The rules in Fig.~\ref{fig:MomRules} are extensions of those for the
number density field $\delta_X(\bm{x})$ given in Fig.~7 of
Ref.~\cite{Mat11}. An important difference of the momentum field from
the density field is the existence of the dotted lines in the top and
bottom graphs. The external, double dotted line corresponds to the
biased momentum, $\bm{j}_X$. The internal, single dotted line
corresponds to the velocity field $\bm{v} = a\dot{\bm{\varPsi}}$. The
internal, wavy line corresponds to the displacement field
$\bm{\varPsi}$. The solid line corresponds to the linear density field
$\delta_\mathrm{L}$. The only one dotted line should be attached to
the external vertex. Numbers of solid and wavy lines attached to the
external vertex are arbitrary, including zero. The index $j$ in the
top graph corresponds to the spatial index of the momentum field
$\bm{j}_X$, and this index appears only in the internal vertex of the
bottom graph. Thus the corresponding factor in the rule for the
external vertex does not depend on the index $j$ and the external
vertex of the top graph only transmits the index to the internal
vertex of the bottom graph. All these properties are the consequences
of the expression of Eq.~(\ref{eq:3-4-01}). For the velocity moment,
we do not consider the redshift-space distortions. In the bottom rule
of Fig.~\ref{fig:MomRules}, we apply an approximation that the LPT
kernel is independent of the time; i.e., perturbations of $n$th-order
$\bm{\varPsi}^{(n)}$ are approximately proportional to $D^n$. If this
approximation is not assumed, $nHf\bar{\bm{L}}_n$ should be replaced
by $nHf\bar{\bm{L}}_n + \partial\bar{\bm{L}}_n/\partial t$.

\begin{figure*}
\begin{center}
\includegraphics[width=32pc]{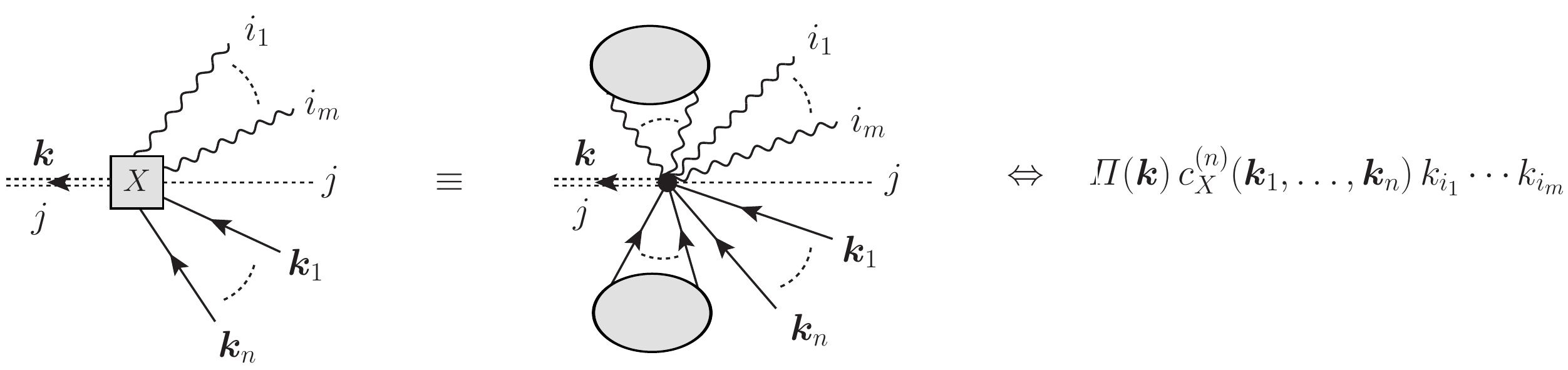}
\caption{\label{fig:ResumMom}
Resummed vertex for the velocity momentum.}
\end{center}
\end{figure*}

The vertex resummation for the velocity momentum can be derived in the
same way as that for the density field in Ref.~\cite{Mat11}. The
diagrammatic rule for the resummed vertex is given in
Fig.~\ref{fig:ResumMom}.
This rule is an extension of that for the density field given in
Fig.~15 of Ref.~\cite{Mat11}. An important difference of the momentum
field from density field is the existence of the dotted line. Only one
dotted line should be attached to the resummed vertex. The grey
ellipse with wavy lines represents the summation of every kind of
diagrams which is attached to the external vertex by any number of
wavy lines. This piece of diagram can be either connected or
disconnected. The same is true for the grey ellipse with solid lines.
Any diagram which is attached to the external vertex by both solid and
wavy lines is not included in the resummation vertex (see
Ref.~\cite{Mat11} for details).

\begin{figure*}
\begin{center}
\includegraphics[width=38pc]{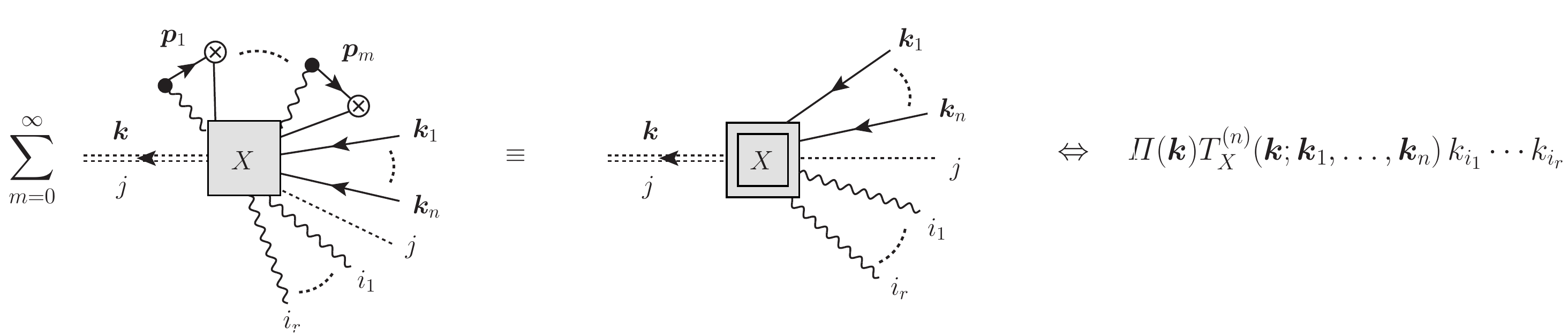}
\caption{\label{fig:SumMomSquare} A further resummation of the
  momentum vertex in the presence of a flat constraint.}
\end{center}
\end{figure*}

\begin{figure*}
\begin{center}
\includegraphics[width=32pc]{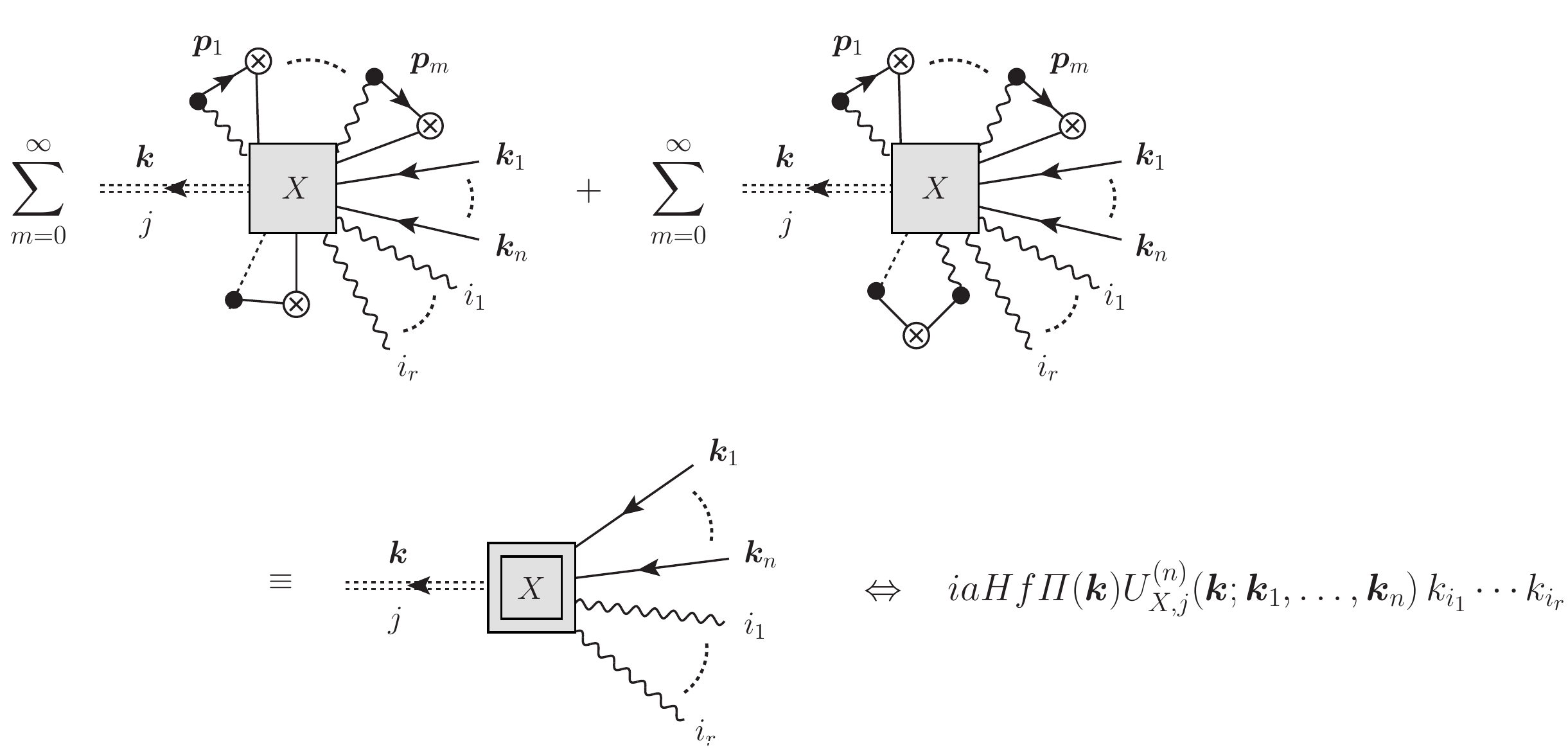}
\caption{\label{fig:SumMomSquare2}
A resummed vertex for the velocity momentum without a dotted line.}
\end{center}
\end{figure*}

A further resummation of the vertex in the presence of a flat
constraint is given in Fig.~\ref{fig:SumMomSquare}.
This resummation is an extension of Fig.~\ref{fig:SumSquare} for the
number density field. The rules are almost the same as those for the
density field, except that they carry the spatial index $j$ for the
momentum $\bm{j}_X$ which should be transmitted to the internal vertex
of the bottom graph in Fig.~\ref{fig:MomRules}. We also define the
double square vertex without a dotted line by
Fig.~\ref{fig:SumMomSquare2}.
This diagram can be calculated by the rules of
Figs.~\ref{fig:SumMomSquare} and \ref{fig:MomRules}. As a result, the
factor of $U^{(n)}_{X,j}$ in Fig.~\ref{fig:SumMomSquare2} is given by
\begin{multline}
  U^{(n)}_{X,j}(\bm{k};\bm{k}_1,\ldots,\bm{k}_n) = 
  \int\frac{d^3p}{(2\pi)^3}
  \bar{L}_{1,j}(-\bm{p}) P_\mathrm{L}(p)
  \\ \times
  \left[
    T^{(n+1)}_X(\bm{k};\bm{k}_1,\ldots\bm{k}_n,\bm{p})
    + T^{(n)}_X(\bm{k};\bm{k}_1,\ldots\bm{k}_n)\,
    \bm{k}\cdot\bar{\bm{L}}_1(\bm{p})
  \right].
  \label{eq:3-4-2}
\end{multline}
The first term is proportional to the derivative $\partial
T^{(n)}_X/\partial k_j$, because we have
\begin{multline}
  T^{(n)}_{X,j}(\bm{k};\bm{k}_1,\ldots,\bm{k}_n)
 \equiv
  \frac{\partial}{\partial k_j}
  T^{(n)}_X(\bm{k};\bm{k}_1,\ldots,\bm{k}_n)
\\
  =
  \int\frac{d^3p}{(2\pi)^3}
  \bar{L}_{n,j}(-\bm{p}) P_\mathrm{L}(p)
  T^{(n+1)}_X(\bm{k};\bm{k}_1,\ldots\bm{k}_n,\bm{p}),
  \label{eq:3-4-3}
\end{multline}
which is straightforwardly shown by substituting the expression of
Eq.~(\ref{eq:3-2-14}). The second term can be calculated by noting
\begin{equation}
  \int\frac{d^3p}{(2\pi)^3}
  \bar{L}_{1,i}(\bm{p}) \bar{L}_{1,j}(-\bm{p}) P_\mathrm{L}(p)
  = -\frac{\delta_{ij}}{3} {\sigma_{-1}}^2.
  \label{eq:3-4-4}
\end{equation}
As a result, Eq.~(\ref{eq:3-4-2}) reduces to
\begin{equation}
  U^{(n)}_{X,j}(\bm{k};\bm{k}_1,\ldots,\bm{k}_n) = 
  \left(
    \frac{\partial}{\partial k_j}
    - \frac{k_j}{3}{\sigma_{-1}}^2
  \right)
  T^{(n)}_X(\bm{k};\bm{k}_1,\ldots\bm{k}_n).
  \label{eq:3-4-5}
\end{equation}

The functions $U^{(n)}_{X,j}$ commonly have an exponential factor, and
we define normalized functions $\hat{U}^{(n)}_{X,j}$ by
\begin{equation}
  U^{(n)}_{X,j}(\bm{k};\bm{k}_1,\ldots,\bm{k}_n) =
  \exp\left(\frac{k^2}{6}\frac{{\sigma_0}^4}{{\sigma_1}^2}\right)\,
  \hat{U}^{(n)}_{X,j}(\bm{k};\bm{k}_1,\ldots,\bm{k}_n).
  \label{eq:3-4-6}
\end{equation}
Substituting Eqs.~(\ref{eq:3-2-17}) and (\ref{eq:3-4-6}) into
Eq.~(\ref{eq:3-4-5}), we have
\begin{equation}
  \hat{U}^{(n)}_{X,j}(\bm{k};\bm{k}_1,\ldots,\bm{k}_n) =
  \left(
    \frac{\partial}{\partial k_j}
    - \frac{k_j}{3} \sigma_\mathrm{dpk}^2
  \right)
  \hat{T}^{(n)}_X(\bm{k};\bm{k}_1,\ldots,\bm{k}_n).
  \label{eq:3-4-7}
\end{equation}
For $n=0,1,2$, Eqs.~(\ref{eq:3-2-18a})--(\ref{eq:3-2-18c}) and
(\ref{eq:3-4-7}) give
\begin{align}
  \hat{U}^{(0)}_{X,j}(\bm{k})
 &= 
  - \frac{k_j}{3} \sigma_\mathrm{dpk}^2,
  \label{eq:3-4-8a}\\
  \hat{U}^{(1)}_{X,j}(\bm{k};\bm{k}_1)
 &=
  - \frac{k_j}{3} \sigma_\mathrm{dpk}^2
    \hat{T}^{(1)}_X(\bm{k};\bm{k}_1)
  - {R_\mathrm{v}}^2 {k_1}^2 \bar{L}_{1,j}(\bm{k}_1),
  \label{eq:3-4-8b}\\
  \hat{U}^{(2)}_{X,j}(\bm{k};\bm{k}_1,\bm{k}_2)
 &=
  - \frac{k_j}{3} \sigma_\mathrm{dpk}^2
    \hat{T}^{(2)}_X(\bm{k};\bm{k}_1,\bm{k}_2)
\nonumber\\
& \qquad
  - {R_\mathrm{v}}^2 {k_1}^2 \bar{L}_{1,j}(\bm{k}_1)
  \hat{T}^{(1)}_X(\bm{k};\bm{k}_2)
\nonumber\\
& \qquad
  - {R_\mathrm{v}}^2 {k_2}^2 \bar{L}_{1,j}(\bm{k}_2)
  \hat{T}^{(1)}_{X}(\bm{k};\bm{k}_1).
  \label{eq:3-4-8c}
\end{align}

Using a somewhat lengthy preparation above, propagators of momentum
field with the flat constraint are straightforwardly obtained by
almost the same method as that in the previous subsection. The
$(n+1)$-point propagator of the momentum field,
$\bm{\varGamma}_X^{\mathrm{v}(n)}(\bm{k}_1,\ldots,\bm{k}_n)$, is
defined by
\begin{multline}
  \left\langle
      \frac{\delta^n \tilde{\bm{j}}_X(\bm{k})}
      {\delta\delta_{\rm L}(\bm{k}_1)
        \cdots\delta\delta_{\rm L}(\bm{k}_n)}
    \right\rangle
   \\ = 
  i\,(2\pi)^{3-3n}\delta_{\rm D}^3(\bm{k}-\bm{k}_{1\cdots n})
  \bm{\varGamma}_X^{\mathrm{v}(n)}(\bm{k}_1,\ldots,\bm{k}_n).
  \label{eq:3-4-9}
\end{multline}

\begin{figure*}
\begin{center}
\includegraphics[width=40pc]{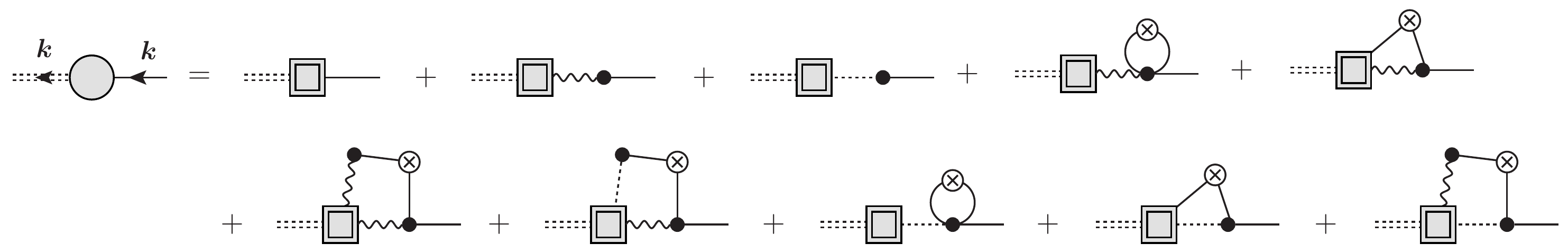}
\caption{\label{fig:PropOneV}
Two-point propagator for the velocity momentum in the one-loop approximation.}
\end{center}
\end{figure*}

The diagrams for the two-point propagator for the momentum field are
given in Fig.~\ref{fig:PropOneV}.
Applying diagrammatic rules of Figs.~\ref{fig:SumMomSquare} and
\ref{fig:SumMomSquare2}, and substituting Eqs.~(\ref{eq:3-2-18a}),
(\ref{eq:3-2-18b}), (\ref{eq:3-4-8a}), and (\ref{eq:3-4-8b}), we
obtain a result,
\begin{multline}
  \frac{\bm{\varGamma}^{\mathrm{v}(1)}_X(\bm{k})}{aHfG_\mathrm{d}(k)} = 
  \left(1-{R_\mathrm{v}}^2 k^2\right)\,
  \bar{\bm{L}}_1(\bm{k})
  \\
  - \frac{\bm{k}}{3} \sigma_\mathrm{dpk}^2 
  \left[
    c^{(1)}_X(k)
    + \left(1 - {R_\mathrm{v}}^2 k^2\right)
    \bm{k}\cdot\bar{\bm{L}}_1(\bm{k})
  \right]
  \\
  + \int\frac{d^3p}{(2\pi)^3} P_\mathrm{L}(p)
  \Biggl\{
    \left(1 - {R_\mathrm{v}}^2 p^2\right)
    \left[
      \bm{k}\cdot\bar{\bm{L}}_2(\bm{k},-\bm{p})
    \right]\,\bar{\bm{L}}_1(\bm{p})
    \\
    + 2
    \left[
      c^{(1)}_X(p)
      + \left(1 - {R_\mathrm{v}}^2 p^2\right)
      \bm{k}\cdot\bar{\bm{L}}_1(\bm{p})
    \right]
    \bar{\bm{L}}_2(\bm{k},-\bm{p})
    \\
    + \frac{3}{2}
    \bar{\bm{L}}_3(\bm{k},\bm{p},-\bm{p})
  \Biggr\}
  + \mbox{[higher-order terms]},
  \label{eq:3-4-10}
\end{multline}
where ``$+$ [higher-order terms]'' indicates the terms which are
proportional to $\sigma_\mathrm{dpk}^2 P_\mathrm{L}$. These terms
correspond to two-loop corrections in the usual sense, and we drop
them in the following. Neglecting mode-coupling terms of
Eq.~(\ref{eq:3-4-10}), the two-point propagator reduces to
\begin{multline}
  \left.
    \bm{\varGamma}_X^{\mathrm{v}(1)}(\bm{k})
  \right|_\mathrm{tree}
  \\
  = aHf\,G_\mathrm{d}(k) 
  \left\{
    b_\mathrm{v}(k)
    - \frac{1}{3} k^2 \sigma_\mathrm{dpk}^2
    \left[ b_\mathrm{v}(k) + c^{(1)}_X(k) \right]
  \right\}
  \frac{\bm{k}}{k^2}
  \\
  = aHf
  \left[
    G_\mathrm{d}(k) b_\mathrm{v}(k)
    - \frac{1}{3} k^2 \sigma_\mathrm{dpk}^2 b^\mathrm{eff}_X(k)
  \right] \frac{\bm{k}}{k^2}.
  \label{eq:3-4-11}
\end{multline}
This result agrees with a previous result, Eq.~(11) of
Ref.~\cite{Bal15} in the peaks model ($\sigma^2_\mathrm{d,pk}$ in their
notation is equal to $\sigma^2_\mathrm{dpk}/3$ in this paper).

\begin{figure*}
\begin{center}
\includegraphics[width=38pc]{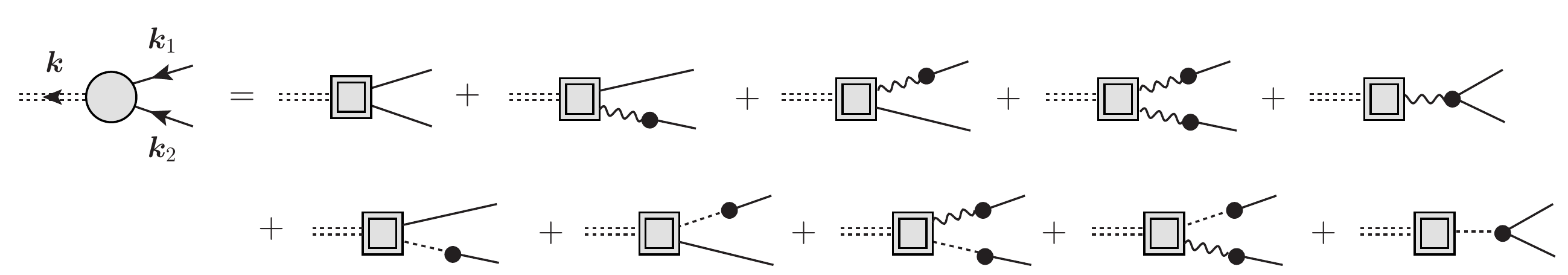}
\caption{\label{fig:PropTwoV}
Three-point propagator for the momentum field in the lowest-order approximation.}
\end{center}
\end{figure*}

The diagrams for the three-point propagator for the momentum field are
given in Fig.~\ref{fig:PropTwoV}.
Applying the diagrammatic rules as before, the result is given by
\begin{multline}
  \frac{\bm{\varGamma}^{\mathrm{v}(2)}_X(\bm{k}_1,\bm{k}_2)}
  {aHfG_\mathrm{d}(k)}
  \\
  =
  \left[
    c^{(1)}_X(k_1) + \left(1 - {R_\mathrm{v}}^2 {k_1}^2\right)
    \bm{k}\cdot\bar{\bm{L}}_1(\bm{k}_1)
  \right]
  \left(1-{R_\mathrm{v}}^2 {k_2}^2\right)\, \bar{\bm{L}}_1(\bm{k}_2)
  \\
  + (\bm{k}_1 \leftrightarrow \bm{k}_2)
  + 2\bar{\bm{L}}_2(\bm{k}_1,\bm{k}_2)
  + \mbox{[higher-order terms]},
  \label{eq:3-4-11-1}
\end{multline}
where $\bm{k}=\bm{k}_1+\bm{k}_2$ and ``$+$ [higher-order terms]''
indicates the terms proportional to $\sigma_\mathrm{dpk}^2$, which
corresponds to one-loop corrections in the usual sense, and we drop them
in the following.

We define the power spectrum of number-weighted velocity field
$P^{\mathrm{v}}_{X,ij}(\bm{k})$ of biased object $X$ by
\begin{equation}
  \left\langle
    j_{X,i}(\bm{k}) j_{X,j}(\bm{k})
  \right\rangle =
  (2\pi)^3 \delta_\mathrm{D}^3(\bm{k}+\bm{k}')
  P^{\mathrm{v}}_{X,ij}(\bm{k}),
  \label{eq:3-4-12}
\end{equation}
where $j_{X,i}$ is the $i$ component of $\bm{j}_X$. Using the
propagators derived above, the number-weighted velocity power spectrum
of biased object $X$ is given by
\begin{multline}
  P^{\mathrm{v}}_{X,ij}(\bm{k}) =
  \varGamma^{\mathrm{v}(1)}_{X,i}(\bm{k})
  \varGamma^{\mathrm{v}(1)}_{X,j}(\bm{k})
  P_\mathrm{L}(k)
  \\
  + \frac{1}{2}
  \int_{\bm{k}_1+\bm{k}_2 = \bm{k}}
  \varGamma^{\mathrm{v}(2)}_{X,i}(\bm{k}_1,\bm{k}_2)
  \varGamma^{\mathrm{v}(2)}_{X,j}(\bm{k}_1,\bm{k}_2)
  P_\mathrm{L}(k_1) P_\mathrm{L}(k_2)
  \\
  + \cdots,
  \label{eq:3-4-13}
\end{multline}
where Gaussian initial conditions are assumed.


\subsection{\label{subsec:DisplaceProp}
Propagators of displacement field with the flat constraint
}

It is also interesting to find statistics of displacement field
$\bm{\varPsi}$ for the biased objects. The number-weighted
displacement field of the biased objects $X$ in Eulerian space can be
defined by
\begin{equation}
  \bm{\psi}^\mathrm{E}_X(\bm{x}) =
  [1 + \delta_X(\bm{x})]
  \bm{\varPsi}^\mathrm{E}(\bm{x}),
  \label{eq:3-5-1}
\end{equation}
where $\bm{\varPsi}^\mathrm{E}(\bm{x})$ is the displacement field in
Eulerian space. The latter is defined by the corresponding value of
the displacement field in Lagrangian space, i.e.,
$\bm{\varPsi}^\mathrm{E}[\bm{x}=\bm{q}+\bm{\varPsi}(\bm{q})] =
\bm{\varPsi}(\bm{q})$. The number-weighted displacement field in
Lagrangian space is defined by
\begin{equation}
  \bm{\psi}^\mathrm{L}_X(\bm{q}) =
  [1 + \delta_X^\mathrm{L}(\bm{q})]
  \bm{\varPsi}(\bm{q}).
  \label{eq:3-5-3}
\end{equation}
The above displacement fields are related by
\begin{equation}
  \bm{\psi}^\mathrm{E}_X(\bm{x}) =
  \int d^3q\, \bm{\psi}_X^\mathrm{L}(\bm{q})\,
  \delta_\mathrm{D}^3[\bm{x}-\bm{q}-\bm{\varPsi}(\bm{q})].
  \label{eq:3-5-4}
\end{equation}

The formalism in the previous subsection to calculate the propagators
of momentum field can be similarly applied to the case of the
number-weighted displacement field. The only difference is the
diagrammatic rule in the bottom graph of Fig.~\ref{fig:MomRules}: the
corresponding factor should be replaced as
$ inaHfL_{n,j} \rightarrow iL_{n,j}$. This replacement is explained by
the fact that the Fourier transform of the number-weighted
displacement field in Eulerian space,
$\tilde{\bm{\psi}}^\mathrm{E}_X$, is just given by replacing
$a\dot{\tilde{\bm{\varPsi}}} \rightarrow \tilde{\bm{\varPsi}}$ in
Eq.~(\ref{eq:3-4-01}). Accordingly, the rules of
Figs.~\ref{fig:ResumMom} and \ref{fig:SumMomSquare} do not change,
and the factor $aHf$ in the rule of Fig.~\ref{fig:SumMomSquare2} is
just removed.

The $(n+1)$-point propagator of the number-weighted displacement
field, $\bm{\varGamma}_X^{\mathrm{d}(n)}(\bm{k}_1,\ldots,\bm{k}_n)$,
is defined by
\begin{multline}
  \left\langle
      \frac{\delta^n \tilde{\bm{\psi}}^\mathrm{E}_X(\bm{k})}
      {\delta\delta_{\rm L}(\bm{k}_1)
        \cdots\delta\delta_{\rm L}(\bm{k}_n)}
    \right\rangle
    \\
    = 
  i\,(2\pi)^{3-3n}\delta_{\rm D}^3(\bm{k}-\bm{k}_{1\cdots n})
  \bm{\varGamma}_X^{\mathrm{d}(n)}(\bm{k}_1,\ldots,\bm{k}_n).
  \label{eq:3-5-5}
\end{multline}

Following the same procedures in the previous subsection, and applying
the above changes, the two-point propagator is given by
\begin{multline}
  \frac{\bm{\varGamma}^{\mathrm{d}(1)}_X(\bm{k})}{G_\mathrm{d}(k)} = 
  \left(1-{R_\mathrm{v}}^2 k^2\right)\,
  \bar{\bm{L}}_1(\bm{k})
  \\
  - \frac{\bm{k}}{3} \sigma_\mathrm{dpk}^2 
  \left[
    c^{(1)}_X(k)
    + \left(1 - {R_\mathrm{v}}^2 k^2\right)
    \bm{k}\cdot\bar{\bm{L}}_1(\bm{k})
  \right]
  \\
  + \int\frac{d^3p}{(2\pi)^3} P_\mathrm{L}(p)
  \Biggl\{
    \left(1 - {R_\mathrm{v}}^2 p^2\right)
    \left[
      \bm{k}\cdot\bar{\bm{L}}_2(\bm{k},-\bm{p})
    \right]\,\bar{\bm{L}}_1(\bm{p})
    \\
    +
    \left[
      c^{(1)}_X(p)
      + \left(1 - {R_\mathrm{v}}^2 p^2\right)
      \bm{k}\cdot\bar{\bm{L}}_1(\bm{p})
    \right]
    \bar{\bm{L}}_2(\bm{k},-\bm{p})
    \\
    + \frac{1}{2}
    \bar{\bm{L}}_3(\bm{k},\bm{p},-\bm{p})
  \Biggr\}
  + \mbox{[higher-order terms]}.
  \label{eq:3-5-6}
\end{multline}
The three-point propagator is given by
\begin{multline}
  \frac{\bm{\varGamma}^{\mathrm{d}(2)}_X(\bm{k}_1,\bm{k}_2)}
  {G_\mathrm{d}(k)}
  \\
  =
  \left[
    c^{(1)}_X(k_1) + \left(1 - {R_\mathrm{v}}^2 {k_1}^2\right)
    \bm{k}\cdot\bar{\bm{L}}_1(\bm{k}_1)
  \right]
  \left(1-{R_\mathrm{v}}^2 {k_2}^2\right)\, \bar{\bm{L}}_1(\bm{k}_2)
  \\
  + (\bm{k}_1 \leftrightarrow \bm{k}_2)
  + \bar{\bm{L}}_2(\bm{k}_1,\bm{k}_2)
  + \mbox{[higher-order terms]},
  \label{eq:3-5-7}
\end{multline}
where $\bm{k}=\bm{k}_1+\bm{k}_2$. The number-weighted displacement
power spectrum is given by a similar expression of
Eq.~(\ref{eq:3-4-13}) with replacements,
$P^\mathrm{v}_{X,ij}(k) \rightarrow P^\mathrm{d}_{X,ij}(k)$ and
$\varGamma^{\mathrm{v}(n)}_{X,i}(k) \rightarrow
\varGamma^{\mathrm{dv}(n)}_{X,i}(k)$.

\section{\label{sec:Conclusion}
  Conclusion}

In this paper, we investigate the effect of velocity bias in the
formalism of iPT. The iPT is a formalism which can evaluate dynamical
evolution of biased objects, based on the Lagrangian perturbation
theory. The peaks model predicts the existence of the velocity bias.
Even though the velocities of peaks and matter are assumed to be the
same at peak locations, the velocities of peaks are statistically
biased with respect to those of matter. With the formalism of iPT, the
dynamical evolution of a biased object and its velocity momentum can be
evaluated by higher-order Lagrangian perturbation theory beyond the
Zel'dovich approximation.

In the first half of this paper, we see how the linear velocity bias
emerges in the framework of iPT. The effects of velocity bias are
already present in the formalism of iPT for the peaks model. Two- and
three-point propagators for the velocity of peaks are calculated. As
an example of their applications, we derive a formula for the one-loop
approximation of velocity dispersion of peaks, which is also biased
with respect to that of matter.

In the second half of this paper, a formal development of iPT in the
presence of velocity bias are presented. We show that the emergence of
velocity bias is a consequence of the flat constraint in general.
Assuming that the flat constraint is the only element with odd parity,
we generally derive the two- and three-point propagators of density in
the one-loop approximation, with a resummation technique regarding the
flat constraint. Diagrammtic rules for the propagators of momentum are
generally derived, and a resummation technique regarding the flat
constraint is also applied.

The formal development presented in this paper gives the basis of
future applications of the iPT formalism. The one-loop power spectra
from the two- and three-point propagators are given by
Eqs.~(\ref{eq:3-3-11}) and (\ref{eq:3-4-13}), and numerical
evaluations of them can be performed by a similar method given in
Ref.~\cite{Mat14}. The purpose of this paper is to provide a
theoretical understanding of the velocity bias with the formalism of
iPT. Numerical evaluations of the power spectra, and comparisons with
numerical simulations, etc., will be addressed in future work.

\begin{acknowledgments}
  The author gratefully thanks Vincent Desjacques and Tobias Baldauf
  for valuable discussions. This work is supported by JSPS KAKENHI
  Grants No.~JP16H03977 and No.~JP19K03835.
\end{acknowledgments}

\appendix
\onecolumngrid

\section{\label{app:IntegralSummation}
Evaluation of the function $T^{(n)}_X$
}

In this appendix, explicit expressions of
$T_X^{(n)}(\bm{k};\bm{k}_1,\ldots,\bm{k}_n)$, defined by
Eq.~(\ref{eq:3-2-14}), are evaluated. With our assumption on the flat
constraint, this function reduces to the form of
Eq.~(\ref{eq:3-2-16}), i.e.,
\begin{equation}
  T_X^{(n)}(\bm{k};\bm{k}_1,\ldots,\bm{k}_n)
  =  \frac{1}{\bar{n}_X}
  \left\langle
    \hat{D}(\bm{k}_1) \cdots \hat{D}(\bm{k}_n)
    \exp\left(- \frac{i}{3} \frac{{\sigma_0}^2}{\sigma_1}
    \bm{k}\cdot\frac{\partial}{\partial\bm{\eta}}
  \right)
  n_X
  \right\rangle.
  \label{eq:a-1}
\end{equation}
To evaluate the above equation, we introduce a vector, $\bm{J} =
{\sigma_0}^2\bm{k}/3\sigma_1$. Using Eqs.~(\ref{eq:3-2-11}) and
(\ref{eq:3-2-13}), the operator which appears in Eq.~(\ref{eq:a-1})
reduces to
\begin{equation}
  \hat{D}(\bm{k}_1)\cdots\hat{D}(\bm{k}_n)
  \exp\left(-i\bm{J}\cdot\frac{\partial}{\partial\bm{\eta}}\right)
  =
  \prod_{a=1}^n
  \left[
    \hat{D}_0(\bm{k}_a) -
    \frac{W(k_aR)}{\sigma_1}\bm{k}_a\cdot\frac{\partial}{\partial\bm{J}}
  \right]
  \exp\left(-i\bm{J}\cdot\frac{\partial}{\partial\bm{\eta}}\right).
  \label{eq:a-2}
\end{equation}
Substituting Eqs.~(\ref{eq:3-1-01}) and (\ref{eq:a-2}) into
(\ref{eq:a-1}), we have
\begin{equation}
  T_X^{(n)}(\bm{k};\bm{k}_1,\ldots,\bm{k}_n)
  =
  \left(\frac{2\pi}{3}\right)^{3/2}
  \frac{1}{\bar{n}_X}
  \left\langle
    \prod_{a=1}^n
    \left[
      \hat{D}_0(\bm{k}_a) -
      \frac{W(k_aR)}{\sigma_1}\bm{k}_a\cdot\frac{\partial}{\partial\bm{J}}
    \right]
    F(\nu,\bm{\zeta},\ldots)
    \exp\left(-i\bm{J}\cdot\frac{\partial}{\partial\bm{\eta}}\right)
    \delta_\mathrm{D}^3(\bm{\eta})
  \right\rangle.
  \label{eq:a-3}
\end{equation}
With our assumption, the variables $\bm{\eta}$ are independent of the
other variables $\nu,\bm{\zeta},\ldots$, and thus the average over
$\bm{\eta}$ is independently evaluated in the above equation. The
probability distribution function of $\bm{\eta}$ is given by
$P(\bm{\eta}) = (3/2\pi)^{3/2} e^{-3\eta^2/2}$. Expressing the delta
function by a Fourier integral, and applying the multidimensional
formula of Gaussian integration, we derive
\begin{align}
  \left\langle
    \exp\left(-i\bm{J}\cdot\frac{\partial}{\partial\bm{\eta}}\right)
    \delta_\mathrm{D}^3(\bm{\eta})
  \right\rangle =
  \left(\frac{3}{2\pi}\right)^{3/2}
  e^{3J^2/2}.
  \label{eq:a-4}
\end{align}
Temporarily putting $\bm{J}=0$ in Eq.~(\ref{eq:a-4}),
Eq.~(\ref{eq:3-1-01}) indicates
\begin{equation}
  \bar{n}_X = \langle F(\nu,\bm{\zeta},\ldots) \rangle.
  \label{eq:a-5}
\end{equation}
Thus we have
\begin{equation}
  T_X^{(n)}(\bm{k};\bm{k}_1,\ldots,\bm{k}_n)
  =
  \frac{1}{\langle F \rangle}
  \left\langle
    \prod_{a=1}^n
    \left[
      \hat{D}_0(\bm{k}_a) -
      \frac{W(k_aR)}{\sigma_1}\bm{k}_a\cdot\frac{\partial}{\partial\bm{J}}
    \right]
    F
  \right\rangle
  e^{3J^2/2}.
  \label{eq:a-6}
\end{equation}
From the above expression, the functions $T_X^{(n)}$ have a common
factor of $\exp(3J^2/2) = \exp(k^2{\sigma_0}^4/6{\sigma_1}^2)$ for every
$n$. We thus define normalized functions $\hat{T}_X^{(n)}$ by
\begin{equation}
  T_X^{(n)}(\bm{k};\bm{k}_1,\ldots,\bm{k}_n) =
  \exp\left(\frac{k^2}{6}\frac{{\sigma_0}^4}{{\sigma_1}^2}\right)\,
  \hat{T}_X^{(n)}(\bm{k};\bm{k}_1,\ldots,\bm{k}_n)
  \label{eq:a-6-1}
\end{equation}

One can evaluate Eq.~(\ref{eq:a-6}) as follows. For $n=0$, we have
$T_X^{(0)}(\bm{k}) = e^{3J^2/2}$, and therefore
\begin{equation}
  \hat{T}_X^{(0)}(\bm{k}) = 1.
  \label{eq:a-7}
\end{equation}
For $n=1$, we have
\begin{equation}
  T_X^{(1)}(\bm{k};\bm{k}_1) =
  \left[
    \frac{\left\langle \hat{D}_0(\bm{k}_1) F \right\rangle}
      {\langle F \rangle}
    - \frac{3}{\sigma_1}
      W(k_1R) \bm{k}_1\cdot\bm{J}
  \right]
  e^{3J^2/2}.
  \label{eq:a-8}
\end{equation}
When $\bm{k}=\bm{0}$ and $\bm{J}=\bm{0}$, we have
$T_X^{(1)}(\bm{0};\bm{k}_1) = c_X^{(1)}(\bm{k}_1)$ from the
definition, Eq.~(\ref{eq:3-2-14}), and the above equation in this case
reduces to
\begin{equation}
  c_X^{(1)}(\bm{k}_1) =
  \frac{\left\langle \hat{D}_0(\bm{k}_1) F \right\rangle}
  {\langle F \rangle}.
  \label{eq:a-9}
\end{equation}
Substituting this equation into Eq.~(\ref{eq:a-8}), the normalized
function is given by
\begin{equation}
  \hat{T}_X^{(1)}(\bm{k};\bm{k}_1) =
  c_X^{(1)}(\bm{k}_1)
  - \frac{{\sigma_0}^2}{{\sigma_1}^2} W(k_1R) \bm{k}\cdot\bm{k}_1.
  \label{eq:a-10}
\end{equation}

For $n=2$, we have
\begin{multline}
  T_X^{(2)}(\bm{k};\bm{k}_1,\bm{k}_2) =
  \left\{
    \frac{\left\langle \hat{D}_0(\bm{k}_1) \hat{D}_0(\bm{k}_2) F \right\rangle}
      {\langle F \rangle}
    - \frac{3}{\sigma_1}
    \left[
      \frac{\left\langle \hat{D}_0(\bm{k}_1) F \right\rangle}
      {\langle F \rangle}
      W(k_2R) \bm{k}_2\cdot\bm{J}
      + (\bm{k}_1 \leftrightarrow \bm{k}_2)
    \right]
    \right.
    \\
    \left.
    + \frac{3}{{\sigma_1}^2} W(k_1R) W(k_2R)
    \left[
      \bm{k}_1\cdot\bm{k}_2
      + 3(\bm{k}_1\cdot\bm{J}) (\bm{k}_2\cdot\bm{J})
    \right]
  \right\}
  e^{3J^2/2}.
  \label{eq:a-11}
\end{multline}
When $\bm{k}=\bm{0}$ and $\bm{J}=\bm{0}$, we have
$T_X^{(2)}(\bm{0};\bm{k}_1,\bm{k}_2) = c_X^{(2)}(\bm{k}_1,\bm{k}_2)$
from the definition, Eq.~(\ref{eq:3-2-14}), and the above equation in
this case reduces to
\begin{equation}
  c_X^{(2)}(\bm{k}_1,\bm{k}_2) =
  \frac{\left\langle \hat{D}_0(\bm{k}_1) \hat{D}_0(\bm{k}_2) F \right\rangle}
  {\langle F \rangle}
  + \frac{3}{{\sigma_1}^2} W(k_1R) W(k_2R) \bm{k}_1\cdot\bm{k}_2.
  \label{eq:a-12}
\end{equation}
Substituting the above equation and Eq.~(\ref{eq:a-9}) into
Eq.~(\ref{eq:a-11}), the normalized function is given by
\begin{equation}
  \hat{T}_X^{(2)}(\bm{k};\bm{k}_1,\bm{k}_2) =
  c_X^{(2)}(\bm{k}_1,\bm{k}_2)
  - \frac{{\sigma_0}^2}{{\sigma_1}^2}
  \left[
    c_X^{(1)}(\bm{k}_1)W(k_2R)\,\bm{k}\cdot\bm{k}_2 +
    (\bm{k}_1 \leftrightarrow \bm{k}_2)
  \right]
  + \frac{{\sigma_0}^4}{{\sigma_1}^4}
  W(k_1R) W(k_2R)
  \left(\bm{k}\cdot\bm{k}_1\right)
  \left(\bm{k}\cdot\bm{k}_2\right).
  \label{eq:a-13}
\end{equation}
The expressions of the function $T_X^{(n)}$ for $n\geq 3$ can be
derived by following the similar procedures above.


\renewcommand{\apj}{Astrophys.~J. }
\newcommand{\aap}{Astron.~Astrophys. }
\newcommand{\aj}{Astron.~J. }
\newcommand{\apjl}{Astrophys.~J.~Lett. }
\newcommand{\apjs}{Astrophys.~J.~Suppl.~Ser. }
\newcommand{\apss}{Astrophys.~Space Sci. }
\newcommand{\jcap}{J.~Cosmol.~Astropart.~Phys. }
\newcommand{\mnras}{Mon.~Not.~R.~Astron.~Soc. }
\newcommand{\mpla}{Mod.~Phys.~Lett.~A }
\newcommand{\pasj}{Publ.~Astron.~Soc.~Japan }
\newcommand{\physrep}{Phys.~Rep. }
\newcommand{\ptp}{Progr.~Theor.~Phys. }
\newcommand{\ptep}{Prog.~Theor.~Exp.~Phys. }
\newcommand{\jetp}{JETP }


\twocolumngrid

\end{document}